\documentclass[11pt,a4paper]{article}
\usepackage{jheppub}
\usepackage{amssymb}
\usepackage{amsfonts}
\usepackage{graphicx}
\usepackage{epstopdf}
\usepackage{dcolumn}
\usepackage{amsmath}
\usepackage{latexsym,bm}
\usepackage{amsthm}
\usepackage{slashed}
\usepackage{float}
\usepackage{extarrows}

\def \be {\begin{equation}}
\def \ee {\end{equation}}
\def \bsp {\begin{split}}
\def \esp {\end{split}}
\def \bea {\begin{eqnarray}}
\def \eea {\end{eqnarray}}

\def\mc{\mathcal}
\def\ptl{\partial}

\def\wtl{\widehat\ptl}

\newcommand{\fracs}[2]{{\textstyle{#1\over #2}}}

\rightline{\today}

\title{Generalized Cartan Calculus in general dimension}

\author[a]{Yi-Nan Wang}

\affiliation[a]{Center for Theoretical Physics,\\Department of Physics\\Massachusetts Institute of Technology\\77 Massachusetts Avenue\\Cambridge, MA 02139, USA}

\emailAdd{wangyn@mit.edu}

\preprint{MIT-CTP/4664}

\abstract{We develop the generalized Cartan Calculus for the groups $G=$SL$(2,\mathbb{R})\times\mathbb{R}^+$, SL$(5,\mathbb{R})$ and SO(5,5). They are the underlying algebraic structures of $d=9,7,6$ exceptional field theory, respectively. These algebraic identities are needed for the ``tensor hierarchy'' structure in exceptional field theory. The validity of Poincar\'{e} lemmas in this new differential geometry is also discussed. Finally we explore some possible extension of the generalized Cartan calculus beyond the exceptional series.}


\begin{document}

\maketitle

\flushbottom

\section{Introduction}

It has long been known that maximally supersymmetric gravity in $d$ dimensions possesses $G=E_{11-d,(11-d)}$ global symmetry\cite{Cremmer:1978}-\cite{deWit:2000wu}. Based on previous works\cite{Koepsell:2000xg}-\cite{HohmKK}, a new paradigm called ``exceptional field theory'' (EFT) was developed in \cite{Exceptional}-\cite{SUSYE6}, to make this hidden symmetry manifest. In this formulation, the space is divided into $d$ ``exterior'' dimensions, and a number of ``interior'' dimensions. The interior part is similar to the extended space in double field theory\cite{Siegel:1993th}-\cite{Hohm:2010xe}, with the dimension higher than the physical dimension. On the extended space, there is a new version of differential geometry. The usual Lie derivative is modified by a term related to certain group invariant tensors. This new Lie derivative is called ``generalized Lie derivative'', denoted by $\mathbb{L}_\Lambda$. Also, the usual Lie bracket is replaced by a corrected ``E-bracket'' $[\cdot,\cdot]_E$. 

Another crucial feature of this extended space is the need for a section condition (or strong constraint). After one solves the section condition, the fields no longer depend on some of the interior dimensions. In double field theory there is a similar story: a $2D$-dimensional extended space is introduced in order to formulate a manifestly T-duality invariant theory. However one half of these dimensions are projected out upon solving the ``strong constraint''. In the exceptional field theory, after solving the section condition, the remaining dimensions along with the exterior dimensions exactly give a 11-dimensional M-theory solution or a 10-dimensional IIB supergravity solution. Hence EFT can be viewed as a unification of M-theory and IIB theory.

In EFT there is Kaluza-Klein 1-form gauge field similar to the 1-form gauge field in the $d$ dimensional maximally supersymmetric gravity. A direct consequence of the modification of differential geometry is that the 2-form field strength associated to this 1-form gauge field is no longer gauge covariant. In order to fix this problem, a 2-form gauge field needs to be introduced and included in the field strength of the 1-form gauge field. Similarly, the 3-form field strength associated with the 2-form gauge field is not gauge covariant, and a 3-form gauge field is required to resolve this. This structure is called ``tensor hierarchy'', first developed in the context of gauged supergravity\cite{deWit:2005hv}\cite{deWit:2008ta}\cite{SamtlebenLec}. The representation of these gauge fields under a group $G$ in EFT is always the same as the corresponding one in the usual maximally supersymmetric gravity.

The derivation of such tensor hierarchy in EFT involves subtle algebraic techniques, especially for cases of large exterior dimension. In \cite{8D}, a natural mathematical structure called ``generalized Cartan calculus'' was developed to simplify these derivations. Analogous to the usual Cartan calculus in differential geometry, the different representations of the gauge fields in EFT are considered as vector spaces of forms. Then a series of projected differential operators $\wtl$, analogous to the exterior derivative $d$, were defined to map one representation into another. Also, a series of binary operators $\bullet$ were constructed to combine two tensor fields in some representation into a tensor field in another representation. These operators enjoy many useful properties. A particularly important formula is called the ``magic formula'':
\be
\mathbb{L}_\Lambda X= \Lambda\bullet\wtl X+\wtl(\Lambda\bullet X),
\ee
acting on some particular tensors $X$. This resembles the Cartan's magic formula in the usual Cartan calculus:
\be
L_\Lambda=i_\Lambda\circ d+d\circ i_\Lambda.
\ee
Also, the $\wtl$ operators are always nil-potent: $\wtl^2=0$. Hence we can define a cohomology structure, called ``exceptional chain complex''. This is analogous to the de Rham cohomology in differential geometry.

The power and magic of the generalized Cartan calculus is that, although it does not contain any information about the exterior dimensions in EFT, one can derive the weight of different tensors in a natural way. When constructing an invariant action of EFT, the number of exterior dimensions is fixed by these weights. Also, with the tools of generalized Cartan calculus, one can derive the tensor hierarchy of gauge fields in an index-free way. This has been done in \cite{8D} and we summarize the related formulas of tensor hierarchy in Section 3. The advantage of the index-free derivation is that it is applicable for EFT with different groups. Once the tensor hierarchy is constructed, and the crucial identities discussed in section 2 hold up to some level, the tensor hierarchy automatically holds to that level. As an example, in this paper the generalized Cartan calculus for $d=9$ theory is explicitly written down. Then the tensor hierarchy of $d=9$ EFT automatically works up to the level of 5-form field strength (one can write down the covariant 2,3,4,5-form field strength, exactly in the same form as the $d=8$ case, see (\ref{fstrengths}) in Section 3).

In this paper, we discuss the relevant generalized Cartan calculus used in $d=9,8,7,6$ EFT. The case $d=9$ is the maximal dimension where an EFT may be constructed, and it has not been discussed in the previous literatures, not even at the level of generalized  Lie derivative and section condition. We give all the necessary formulas explicitly. We also analyze the validity of a Poincar\'{e} lemma (or local exactness) associated to the exceptional chain complex. Generally a local exactness statement only holds in one particular solution to the section condition, but not in the other solution. This is also an unusual feature of this new differential geometry. 

The rest of this paper is organized in the following way: in section 2 we discuss the general structure of generalized Cartan Calculus that can be used in an arbitrary EFT. In section 3 we review the basics of tensor hierarchy in EFT and how to use the generalized Cartan Calculus in deriving it. In section 4, 5, 6, 7, the explicit formula for SL$(2,\mathbb{R})\times \mathbb{R}_+$, SL$(2,\mathbb{R})\times $SL$(3,\mathbb{R})$, SL$(5,\mathbb{R})$ and SO(5,5) generalized Cartan calculus are presented. They correspond to $d=9$, $d=8$, $d=7$ and $d=6$ EFTs, respectively(both of these equivalent notations are used in different contexts). In section 8, we discuss the generalized Cartan Calculus for other groups. For the $E_{6(6)}$ theory which corresponds to $d=5$ EFT and $O(D,D)$ theory which corresponds to double field theory, the generalized Cartan calculus is quite trivial. We also examine the extension to other groups and representations. We conclude with an outlook in section 9.

\section{General structure of generalized Cartan Calculus}

\subsection{Generalized Lie derivative and vectors}

To construct a generalized Cartan Calculus, one first picks a symmetry group $G$, and a representation $R$ of $G$ in which the vector field $V^M$ lives. The dimension of the interior space is equal to the dimension of representation dim$(R)$. To make the theory fully covariant under the symmetry group $G$, all the tensors appearing in the theory should be $G$-covariant tensors.

The usual notion of $d$-dimensional EFT corresponds to a EFT with $d$ exterior coordinates and symmetry group $G=E_{11-d,(11-d)}$ ($E_{2(2)}=$SL$(2,\mathbb{R})\times \mathbb{R}_+$, $E_{3(3)}=$SL$(2,\mathbb{R})\times $SL$(3,\mathbb{R})$, $E_{4(4)}=$SL$(5,\mathbb{R})$, $E_{5(5)}=$SO(5,5)). The representations $R$ of the vector fields and interior coordinates are given in Table \ref{t:grouprep}. Throughout the paper we use $\mu,\nu,\dots$ to denote the exterior directions.

\begin{table}
\centering
\begin{tabular}{cccc}
\hline
$d$&Lie group $G$&rep $R$&$D$\\
\hline
9&SL$(2,\mathbb{R})\times \mathbb{R}_+$&$\mathbf{2+1}$&3\\
8&SL$(3,\mathbb{R})\times $SL$(2,\mathbb{R})$&$(\mathbf{3},\mathbf{2})$&6\\
7&SL$(5,\mathbb{R})$&$\mathbf{10}$(antisymmetric rank-2 tensor)&10\\
6&SO$(5,5)$&$\mathbf{16}$(Majorana-Weyl spinor)&16\\
5&$E_{6(6)}$&$\mathbf{27}$&27\\
4&$E_{7(7)}$&$\mathbf{56}$&56\\
3&$E_{8(8)}$&-&-\\
\hline
\end{tabular}
\caption{The symmetry group $G$, representation of 1-form field $R$ and its dimension $D$ of the $d$-dimensional EFT. The representation $R$ corresponds to the representation of 1-form gauge field in $d$-dimensional EFT, and $D$ is equal to the dimension of interior space $M_i$. They are exactly the same as the global symmetry group and the representation of vector fields of maximally supersymmetric gravity in $d$-dimensions.}\label{t:grouprep}
\end{table}

The ordinary (interior) Lie derivative of a vector,
\be
L_\Lambda V^M=\Lambda^N\ptl_N V^M-V^N\ptl_N\Lambda^M,
\ee
is modified to a new form:
\be
\mathbb{L}_\Lambda V^M=\Lambda^N\ptl_N V^M+\alpha\mathbb{P}^M{}_N{}^K{}_L{}\ptl_K\Lambda^L V^N+\lambda\ptl_N\Lambda^N V^M.\label{gdv}
\ee
The latter one is called ``generalized Lie derivative'', $\mathbb{P}^M{}_N{}^K{}_L{}$ is the projector to the adjoint representation of $G$, which is a $G$-invariant tensor\footnote{For the cases in which $G$ is not semi-simple, this projector term is a linear combination of the projectors to the adjoint of each group factor. For example in the case $d=8$, it is $ -2  (\mathbb{P}_{(8,1)})^{M}{}_{N}{}^{P}{}_{Q}\partial_P\Lambda^Q V^N
   -3 (\mathbb{P}_{(1,3)})^{M}{}_{N}{}^{P}{}_{Q}\partial_P\Lambda^Q V^N$}.
The parameter $\lambda$ defines the density weight of the vector.

The usual Lie bracket of two vectors is also modified to the following E-bracket:
\be
[U,V]_E^M=U^N\ptl_N V^M+\fracs{1}{2}Z^{MN}{}_{PQ}\ptl_N U^P V^Q-(U\leftrightarrow V).
\ee
$Z^{MN}{}_{PQ}$ is also a $G$-invariant tensor. In this paper all the $Z$-tensors are symmetric in $MN$ and $PQ$:
\be
Z^{MN}{}_{PQ}=Z^{NM}{}_{PQ}=Z^{MN}{}_{QP}=Z^{NM}{}_{QP}.
\ee

With this $Z$-tensor, the generalized Lie derivative can be rewritten in the following form:
\be
\mathbb{L}_\Lambda V^M=\Lambda^N\ptl_N V^M-V^N\ptl_N\Lambda^M+Z^{MN}{}_{PQ}\ptl_N \Lambda^P V^Q+(\lambda-\omega)\ptl_N\Lambda^N V^M.\label{gdvomega}
\ee
For each EFT, there is a distinguished weight $\omega$, such that the last term above vanishes. From the definition we have the following identity for vectors $U,V$ of weight $\omega$:
\be
[U,V]_E=\fracs{1}{2}(\mathbb{L}_U V-\mathbb{L}_V U).\label{antisym}
\ee
In the following discussions all the gauge parameters $\Lambda$ in generalized Lie derivative $\mathbb{L}_\Lambda$ are assumed to have weight $\omega$.

A consistency requirement is the algebraic closure condition:
\be
[\mathbb{L}_U,\mathbb{L}_V]W^M=\mathbb{L}_{[U,V]_E}W^M.\label{closure}
\ee

To guarantee the closure constraint (\ref{closure}), we impose the following ``section condition'':
\be
Z^{MN}{}_{PQ}\ptl_M\otimes\ptl_N=0.\label{section}
\ee
Apart from that, the following condition needs to be satisfied for (\ref{closure}) to hold:
\be
Z^{M(L|}{}_{PQ}Z^{P|N)}{}_{RS}=Z^{M(N|}{}_{RS}\delta^{|L)}_Q\label{ZZ}.
\ee
Then the value of $\alpha$ in (\ref{gdv}) can be fixed. For $5\leq d\leq 8$, $\alpha$ and the expression of $Z^{MN}{}_{PQ}$  can be found in \cite{Berman:2012vc}. 

\subsection{Higher tensor representations and generalized Cartan calculus}

As comprehensively discussed for the $d=8$ case in \cite{8D}, there are other types of tensors, in addition to vectors $V^M$,  carrying different weights. 

We denote the set of vectors with weight $\lambda$ by $\mathfrak{A}(\lambda)$. Then there are sets of tensors $\mathfrak{B}(\lambda)$, $\mathfrak{C}(\lambda)$, $\mathfrak{D}(\lambda)$, $\mathfrak{E}(\lambda)$, $\mathfrak{F}(\lambda)$, etc. We use the notation $\mathfrak{X}_n(\lambda)$ for a general tensor with weight $\lambda$, for instance, $\mathfrak{X}_2(\lambda)\equiv\mathfrak{B}(\lambda)$. In EFT the $n$-form gauge fields are elements of $\mathfrak{X}_n(n\omega)$, where $\omega$ is the distinguished weight mentioned before. Its representation should coincide with the representation of $n$-form gauge field in the maximally supersymmetric gravity in the corresponding dimension.  For the dual of $n$-form gauge field in $d$ dimensions, that is, a $(d-n-2)$-form gauge field, it is in the contragredient representation of $n$-form gauge field. We write them down explicitly in Table \ref{t:reps}, for $5\leq d\leq 9$.

\begin{table}
\centering
\begin{tabular}{cccccccc}
\hline
$d$&Lie group $G$&1-form/$\mathfrak{A}$&2-form/$\mathfrak{B}$&3-form/$\mathfrak{C}$&4-form/$\mathfrak{D}$&5-form/$\mathfrak{E}$&6-form/$\mathfrak{F}$\\
\hline
9&$SL(2,R)\times R_+$&$\mathbf{2+1}$&$\mathbf{2}$&$\mathbf{1}$&$\mathbf{1}$&$\mathbf{2}$&$\mathbf{2+1}$\\
8&$SL(3,R)\times SL(2,R)$&$(\mathbf{3},\mathbf{2})$&$(\mathbf{\bar{3}},\mathbf{1})$&$(\mathbf{1},\mathbf{2})$&$(\mathbf{3},\mathbf{1})$&$(\mathbf{\bar{3}},\mathbf{2})$&-\\
7&$SL(5,R)$&$\mathbf{10}$&$\mathbf{\bar{5}}$&$\mathbf{5}$&$\mathbf{\overline{10}}$&-&-\\
6&$SO(5,5)$&$\mathbf{16}$&$\mathbf{10}$&$\mathbf{\overline{16}}$&-&-&-\\
5&$E_{6(6)}$&$\mathbf{27}$&$\mathbf{\overline{27}}$&-&-&-&-\\
\hline
\end{tabular}
\centering
\caption{Representation of tensor objects and the corresponding $n$-form gauge fields in EFT for $5\leq d\leq 9$. For the group SL(2,R), the contragredient (dual) of a representation is itself.}\label{t:reps}
\end{table}

There is also a set of projected differential operators denoted by $\widehat\ptl $, acting differently on different tensor spaces\footnote{This differential operator was first introduced in \cite{Cederwall:2013naa}. The author thanks Martin Cederwall for pointing out that.}. This operator is nil-potent: $\widehat\ptl ^2=0$, guaranteed by the section condition. We have the following chain complex:
\be
\mathfrak{A}(\omega)\xlongleftarrow{\widehat\ptl }\mathfrak{B}(2\omega)\xlongleftarrow{\widehat\ptl }\mathfrak{C}(3\omega)\xlongleftarrow{\widehat\ptl }\mathfrak{D}(4\omega)
\xlongleftarrow{\widehat\ptl }\mathfrak{E}(5\omega)\dots\xlongleftarrow{\widehat\ptl}\mathfrak{X}_l(l\omega)
\ee
We denote the set of highest tensor objects in the theory by $\mathfrak{X}_l(l\omega)$. And we define the length of exceptional chain complex to be $l$. For the generalized Cartan Calculus of $d$ dimensional EFT, $l=d-3$.

One can now define the cohomology analogous to the de Rham cohomology. But as discussed later in specific cases, the Poincar\'{e} lemma does not always hold, thus its geometric meaning is not clear.

The quantity $\wtl B$ ($B\in\mathfrak{B}(2\omega)$) should always be a ``trivial gauge parameter'', that is, $\mathbb{L}_{\wtl B}$ always gives 0. A way of realizing this condition is to write 
\be
(\wtl B)^M\propto Z^{MN}{}_{PQ}\ptl_N B^{PQ},\label{wtlB}
\ee
then $\mathbb{L}_{\wtl B}=0$ is a consequence of the identity (\ref{ZZ}). In general there will be an isomorphism that transform $B^{PQ}$ to $B(2\omega)$, and vice versa:
\be
\bsp
&I_B:\ (R\times R)_{\text{sym}}\rightarrow B(2\omega):\ I_B(B^{PQ})=B\\
&I^{-1}_B:\ B(2\omega)\rightarrow(R\times R)_{\text{sym}}:\ I^{-1}_B(B)=B^{PQ}.\label{PRB}
\end{split}
\ee
Then $(\wtl B)^M$ acting on the tensor $B$ can be defined as:
\be
(\wtl B)^M\propto Z^{MN}{}_{PQ}\ptl_N (I^{-1}_{B} B)^{PQ}.
\ee

In addition to $\widehat\ptl $, there is a set of binary operators
\be
\bullet:(\mathfrak{X}_a(\omega_a),\mathfrak{X}_b(\omega_b))\rightarrow \mathfrak{X}_{a+b}(\omega_a+\omega_b).
\ee
They can be viewed as projectors to some particular representation among the representations given by tensor product. For example, for $d=8$, $a=1$, $b=1$, the tensor product rule is:
\be
(\mathbf{3},\mathbf{2})\times(\mathbf{3},\mathbf{2})=(\mathbf{\bar{3}},\mathbf{1})+(\mathbf{\bar{3}},\mathbf{3})+(\mathbf{6},\mathbf{1})+(\mathbf{6},\mathbf{3}),
\ee
and the $\bullet$ projector only gives the representation in $(\mathbf{\bar{3}},\mathbf{1})$.

Although we use this single universal notation $\bullet$, the rule of $\bullet$ acting on different type of tensors are actually different. If $a\neq b$, that is, $X\in\mathfrak{X}_a$ and $Y\in\mathfrak{X}_b$ are different types of tensors, we always define that
\be
X\bullet Y\equiv Y\bullet X.
\ee
For the case $a=b=1$, we can explicitly write out the rule of $\bullet$ using the isomorphism $I_B$ defined in (\ref{PRB}): for $A_1,A_2\in\mathfrak{A}(\omega)$,
\be
(A_1\bullet A_2)\propto I_B(A_{(1}^P A_{2)}^Q).
\ee
From this, we can see that the operator $\bullet$ acting on $A_1,A_2\in\mathfrak{A}$ is commutative:
\be
A_1\bullet A_2=A_2\bullet A_1.
\ee
Then from (\ref{gdvomega}) and (\ref{wtlB}), there is the following identity: for $A_1,A_2\in\mathfrak{A}(\omega)$,
\be
\mathbb{L}_{A_1}A_2+\mathbb{L}_{A_2}A_1=\widehat\ptl (A_1\bullet A_2)\label{AAprop}.
\ee 
Also there is the following Jacobi identity:
\be
[[A_1,A_2]_E,A_3]_E+\mathrm{cycl.}\ =\ \fracs{1}{6}([A_1,A_2]_E\bullet A_3)+\mathrm{cycl.},
\ee
or equivalently
\be
\mathbb{L}_{A_{[1}}\mathbb{L}_{A_2} A_{3]}\ =\ \fracs{1}{3}([A_{[1},A_2]_E\bullet A_{3]}).\label{Jacobi2}
\ee

For the case $a+b=l+1=d-2$, the two tensors involved are mutually contragredient, and the $\bullet$ operator results in a singlet. Usually it is also a scalar with weight 1, which means $\omega=1/(d-2)$, as confirmed in specific dimensions (this also corresponds to the quantity $\beta_{(11-d)}$ in \cite{Berman:2012vc}). Indeed a scalar can act as Lagrangian density only if it has weight 1, because the generalized Lie derivative acting on the action would be:
\be
\mathbb{L}_\Lambda\int L=\int(\Lambda^M\ptl_M L+\ptl_M\Lambda^M L)=\int\ptl_M(\Lambda^M L)=0,
\ee
when the boundary term is ignored.

One can define the way how $\mathbb{L}_\Lambda$ acts on different tensors. The general requirement is, for any two tensors $X$, $Y$ listed above, the distribution law holds:
\be
\mathbb{L}_\Lambda(X\bullet Y)=X\bullet\mathbb{L}_\Lambda Y+\mathbb{L}_\Lambda X\bullet Y.\label{distribute}
\ee

\subsection{Magic formulas and other identities}

There is a set of additional identities which make the gauge hierarchy of EFT work to the level of 5-form field strength.

The first class of identities is called ``magic formulas'':

For $\Lambda\in\mathfrak{A}(\omega),X\in\mathfrak{X}_n(n\omega)$, $l>n>1$
\be
\mathbb{L}_\Lambda X=\Lambda\bullet\widehat\ptl  X+\widehat\ptl (\Lambda\bullet X).\label{magic}
\ee

With these identities, one can easily prove that the generalized Lie derivative always commutes with projected differential operator $\widehat\ptl $, for $X\in\mathfrak{X}_n(n\omega)\ ,2\leq n<l$:
\be
\mathbb{L}_\Lambda(\widehat\ptl X)=\widehat\ptl (\mathbb{L}_\Lambda X).
\ee
For example, let us prove this for any $B\in\mathfrak{B}(2\omega)$,
\be
\mathbb{L}_\Lambda\widehat\ptl B=\widehat\ptl(\mathbb{L}_\Lambda B).
\ee
We use the formula (\ref{AAprop}) with $A_1=\widehat\ptl B,A_2=\Lambda$, and recall that $\wtl B$ is a trivial gauge parameter, then the l.h.s. above gives:
\be
\mathbb{L}_\Lambda\widehat\ptl B=\widehat\ptl(\Lambda\bullet\wtl B)-\mathbb{L}_{\wtl B}\Lambda=\widehat\ptl(\Lambda\bullet\wtl B).
\ee
For the r.h.s., we use the magic formula (\ref{magic}) with $X=B$, 
\be
\wtl(\mathbb{L}_\Lambda B)=\wtl(\Lambda\bullet\wtl B+\wtl(\Lambda\bullet B))=\widehat\ptl(\Lambda\bullet\wtl B).
\ee
Similarly for $C\in\mathfrak{C}(3\omega)$, use the magic formula with $X=\wtl C$ and $X=C$ on l.h.s and r.h.s. respectively, 
\be
\mathbb{L}_\Lambda\widehat\ptl C=\Lambda\bullet\wtl^2 C+\widehat\ptl(\Lambda\bullet\wtl C)=\wtl(\Lambda\bullet\wtl C),
\ee
\be
\wtl(\mathbb{L}_\Lambda C)=\wtl(\Lambda\bullet\wtl C+\wtl(\Lambda\bullet C))=\widehat\ptl(\Lambda\bullet\wtl C).\label{wtlC}
\ee
Hence $\mathbb{L}_\Lambda\widehat\ptl C=\widehat\ptl(\mathbb{L}_\Lambda C)$.

One can see that they are direct consequences of the nil-potent property of projected differential operator $\wtl$.

For $n=l$, one may also check that generalized Lie derivative commutes with $\widehat\ptl $, so that the following diagram commutes:
\be
\begin{array}{ccccccccccc}\mathfrak{X}_1(\omega)&\xlongleftarrow{\hat\partial}&\mathfrak{X}_2(2\omega)&\xlongleftarrow{\hat\partial}& \cdots&
\xlongleftarrow{\hat\partial}&\mathfrak{X}_{l-1}((l-1)\omega)&\xlongleftarrow{\hat\partial}&\mathfrak{X}_l(l\omega)\\
\Big{\downarrow}{\mathbb{L}_\Lambda}&&\Big{\downarrow}{\mathbb{L}_\Lambda}&& &&\Big{\downarrow}{\mathbb{L}_\Lambda}&&\Big{\downarrow}{\mathbb{L}_\Lambda}\\
\mathfrak{X}_1(\omega)&\xlongleftarrow{\hat\partial}&\mathfrak{X}_2(2\omega)&\xlongleftarrow{\hat\partial}&\cdots&
\xlongleftarrow{\hat\partial}&\mathfrak{X}_{l-1}((l-1)\omega)&\xlongleftarrow{\hat\partial}&\mathfrak{X}_l(l\omega)
\end{array}
\ee

Using these identities, it is also possible to prove the gauge closure identity for higher tensors in an index-free way. We want to prove that for $X\in\mathfrak{X}_n(n\omega)$, $l>n>1$,
\be
\mathbb{L}_U\mathbb{L}_V X-\mathbb{L}_V\mathbb{L}_U X=\mathbb{L}_{[U,V]_E}X.
\ee
We write the term $\mathbb{L}_U\mathbb{L}_V X$ in two different ways:
\be
\bsp
\mathbb{L}_U\mathbb{L}_V X\ =\ &\mathbb{L}_U(V\bullet\wtl X)+\mathbb{L}_U\wtl(V\bullet X)\\[0.5ex]
\ =\ &\mathbb{L}_U V\bullet\wtl X+V\bullet\mathbb{L}_U\wtl X+\wtl(\mathbb{L}_U V\bullet X)+\wtl(V\bullet\mathbb{L}_U X),
\end{split}
\ee
where we rewrite $\mathbb{L}_V X$ using the magic formula (\ref{magic}), the distribution property (\ref{distribute}), and the fact that $\mathbb{L}_U$ commutes with $\wtl$.
\be
\bsp
\mathbb{L}_U\mathbb{L}_V X\ =\ &U\bullet\wtl(\mathbb{L}_V X)+\wtl(U\bullet\mathbb{L}_V X)\\[0.5ex]
\ =\ &U\bullet\mathbb{L}_V\wtl X+\wtl(U\bullet\mathbb{L}_V X),
\end{split}
\ee
where we insert $\mathbb{L}_V X$ as $X$ in the magic formula (\ref{magic}).

Adding these two parts, we arrived at
\be
\bsp 
2\mathbb{L}_U\mathbb{L}_V X-(U\leftrightarrow V)=\mathbb{L}_U V\bullet\wtl X+\wtl(\mathbb{L}_U V\bullet X)-(U\leftrightarrow V).\label{LULV}
\end{split}
\ee
Then we write
\be
\bsp
2\mathbb{L}_{[U,V]_E}X\ =\ &\mathbb{L}_{\mathbb{L}_U V-\mathbb{L}_V U}X\\[0.5ex]
\ =\ &\mathbb{L}_U V\bullet\wtl X+\wtl(\mathbb{L}_U V\bullet X)-(U\leftrightarrow V),
\end{split}
\ee
where we used the antisymmetrization property (\ref{antisym}) and the magic formula as well.

Comparing with (\ref{LULV}) we conclude that
\be
\mathbb{L}_U\mathbb{L}_V X-\mathbb{L}_V\mathbb{L}_U X=\mathbb{L}_{[U,V]_E}X.
\ee

The second class of identities involves both $\widehat\ptl $ and $\bullet$ but no $\mathbb{L}_\Lambda$:

(1) For any $B_1,B_2\in\mathfrak{B}(2\omega)$,
\be
\widehat\ptl  B_1\bullet B_2-\widehat\ptl  B_2\bullet B_1=\widehat\ptl (B_1\bullet B_2)\label{BBprop}.
\ee

(2) For any $B\in\mathfrak{B}(2\omega),C\in\mathfrak{C}(3\omega)$,
\be
\widehat\ptl  B\bullet C+B\bullet\widehat\ptl  C=\widehat\ptl (B\bullet C)\label{BCprop}.
\ee

The third class of identities only involves $\bullet$:

(1) When $\bullet$ acts on $B_1,B_2\in\mathfrak{B}$, it is anti-commutative:
\bea
B_1\bullet B_2&=&-B_2\bullet B_1.
\eea

(2) For any $A_1,A_2,A_3\in\mathfrak{A}$,
\be
A_1\bullet (A_2\bullet A_3)+A_2\bullet (A_3\bullet A_1)+A_3\bullet (A_1\bullet A_2)=0\label{AAAprop}.
\ee

(3)
For any $A_1,A_2\in\mathfrak{A},B\in\mathfrak{B}$,
\be
A_1\bullet(A_2\bullet B)+A_2\bullet (A_1\bullet B)+B\bullet(A_1\bullet A_2)\ =\ 0 \;. \label{AABprop}
\ee

For a specific group $G$, these additional algebraic identities may not all hold. For the group $G$ that corresponds to $d$ dimensional EFT, an identity holds if and only if all the $\bullet$ operators involved result in a tensor with weight less than $(l+1)\omega=1$.

These set of rules in generalized Cartan calculus exactly imply that the tensor hierarchy of EFT holds up to $(d-3)$-form field strength. One can derive the whole gauge structure in an index-free and universal way. For example, in $d=8$ EFT\cite{8D}, one can explicitly construct a series of $n$-form gauge fields which are tensors in $\mathfrak{X}_n(n/6)$, respectively, and a series of $n+1$-form gauge covariant field strengths, up to $n=4$. 

We classify the necessary identities used at each level of the tensor hierarchy. The existence of $n$-form gauge covariant field strengths requires the existence of the lower form gauge covariant field strengths. The notations $A,B,C,D$ are defined in the same way as before.

\begin{itemize}
\item{The existence of 2-form gauge covariant field strength $\mc{F}_{\mu\nu}$ requires: 
\be
\mathbb{L}_{A_1}A_2+\mathbb{L}_{A_2}A_1=\widehat\ptl (A_1\bullet A_2).\label{Forms2}
\ee
}
\item{The existence of 3-form gauge covariant field strength $\mc{H}_{\mu\nu\rho}$ requires: 
\be
\bsp
&\mathbb{L}_A B=A\bullet\widehat\ptl  B+\widehat\ptl (A\bullet B).
\label{Forms3}
\end{split}
\ee
}
\item{The existence of 4-form gauge covariant field strength $\mc{J}_{\mu\nu\rho\sigma}$ requires: 
\be
\bsp
&\mathbb{L}_A C=A\bullet\widehat\ptl  C+\widehat\ptl (A\bullet C).\\[0.5ex]
&\widehat\ptl  B_1\bullet B_2-\widehat\ptl  B_2\bullet B_1=\widehat\ptl (B_1\bullet B_2).\\[0.5ex]
&A_1\bullet (A_2\bullet A_3)+A_2\bullet (A_3\bullet A_1)+A_3\bullet (A_1\bullet A_2)=0.\label{Forms4}
\end{split}
\ee
}
\item{The existence of 5-form gauge covariant field strength $\mc{K}_{\mu\nu\rho\sigma\tau}$ requires: 
\be
\bsp
&\mathbb{L}_A D=A\bullet\widehat\ptl D+\widehat\ptl (A\bullet D).\\[0.5ex]
&\widehat\ptl  B\bullet C+B\bullet\widehat\ptl  C=\widehat\ptl (B\bullet C).\\[0.5ex]
&A_1\bullet(A_2\bullet B)+A_2\bullet (A_1\bullet B)+B\bullet(A_1\bullet A_2)\ =\ 0 \;.\\[0.5ex]
&B_1\bullet B_2=-B_2\bullet B_1\label{Forms5}
\end{split}
\ee
}
\end{itemize}

Finally we comment briefly on the necessity of the strong form of section condition (\ref{section}). Analogous to the situation in double field theory, the strong form of section condition means that all the products $Z^{MN}{}_{PQ}\ptl_M A\ptl_N B$ vanish. This is equivalent to the condition that $Z^{MN}{}_{PQ}\ptl_M\ptl_N A$ vanishes for all the products of fields and gauge parameters $A$. The weak form of the section condition only requires that $Z^{MN}{}_{PQ}\ptl_M\ptl_N A$ vanishes, where $A$ is a single field or gauge parameter. The nil-potent property of $\wtl$ only requires the weak version. All the identities listed before that only involve a single derivative also holds (for example the magic formulas), as there is no need for section condition at all. However, if only the weak section condition holds, $\mathbb{L}_\Lambda$ generally does not commutes with $\wtl$, because in the derivation (\ref{wtlC}) we used $\wtl^2(\Lambda\bullet C)=0$, which is a consequence of strong section condition. Consequently, the closure constraint only holds subject to the strong constraint.
 
\section{Tensor Hierarchy in EFT}

In this section we briefly summarize the tensor hierarchy in EFT up to the gauge covariant 5-form field strength developed in \cite{8D}. 

First we have the Kaluza Klein 1-form field $A_\mu\in\mathfrak{A}(\omega)$, and the covariant derivative is defined as:
\be
\mc{D}_\mu=\ptl_\mu-\mathbb{L}_{A_\mu}.
\ee
The gauge transformation of $A_\mu$ is defined as
\be
\delta_\Lambda A_\mu^M=\mc{D}_\mu\Lambda\label{variA}.
\ee
Gauge covariance of a tensor $X$ means that $\delta_\Lambda X=\mathbb{L}_\Lambda X$. One can check that for a gauge covariant $X$, $\mc{D}_\mu X$ is also gauge covariant. Another straightforward property is that $\mc{D}_\mu$ commutes with $\wtl$, because $\mathbb{L}_{A_\mu}$ does.

Now we want to construct a gauge covariant 2-form field strength, that will appear in the action. The naive field strength is an analogue of field strength in Yang-Mills theory, which replaces the Lie algebra bracket by the E-bracket:
\be
 F_{\mu\nu}{} \ = \ 2\ptl_{[\mu}A_{\nu]}-[A_\mu,A_\nu]_E.
\ee
Now the general variation of $F_{\mu\nu}$ is
\be
\bsp
\delta F_{\mu\nu}\ =\ &2\ptl_{[\mu}\delta A_{\nu]}-2[A_{[\mu},\delta A_{\nu]}]_E\\[0.5ex]
\ =\ &2\ptl_{[\mu}\delta A_{\nu]}-\mathbb{L}_{A_{[\mu}}\delta A_{\nu]}+\mathbb{L}_{\delta A_{[\nu}}A_{\mu]}\\[0.5ex]
\ =\ &2\ptl_{[\mu}\delta A_{\nu]}-2\mathbb{L}_{A_{[\mu}}\delta A_{\nu]}+\wtl(A_{[\mu}\bullet\delta A_{\nu]})\\[0.5ex]
\ =\ &2\mc{D}_{[\mu}\delta A_{\nu]}+\wtl(A_{[\mu}\bullet\delta A_{\nu]}).
\end{split}
\ee
From the first line to second line we used (\ref{antisym}), and from the second line to third line we used (\ref{AAprop}). Now insert (\ref{variA}), and use the following identity
\be
[\mc{D}_\mu,\mc{D}_\nu]_E=-\mathbb{L}_{F_{\mu\nu}},
\ee
We find that the failure of gauge covariance is
\be
\delta_\Lambda F_{\mu\nu}-\mathbb{L}_\Lambda F_{\mu\nu}=\wtl(A_{[\mu}\bullet\mc{D}_{\nu]}\Lambda-\Lambda\bullet F_{\mu\nu}).
\ee
To resolve this problem, we introduce the 2-form gauge field $B_{\mu\nu}\in\mathfrak{B}(2\omega)$, and modify $F_{\mu\nu}$ to a corrected field strength:
\be
\mc{F}_{\mu\nu}=F_{\mu\nu}+\wtl B_{\mu\nu}.
\ee
The general variation of $\mc{F}_{\mu\nu}$ can be rewritten as:
\be
\bsp
\delta\mc{F}_{\mu\nu}\ =\ &2\mc{D}_{[\mu}\delta A_{\nu]}+\wtl(\delta B_{\mu\nu}+A_{[\mu}\bullet\delta A_{\nu]})\\[0.5ex]
\ =\ &2\mc{D}_{[\mu}\delta A_{\nu]}+\wtl(\Delta B_{\mu\nu}),
\end{split}
\ee
where we defined
\be
\Delta B_{\mu\nu}=\delta B_{\mu\nu}+A_{[\mu}\bullet\delta A_{\nu]}.
\ee
Then if $\Delta_\Lambda B_{\mu\nu}=\Lambda\bullet \mc{F}_{\mu\nu}$, we have
\be
\delta_\Lambda\mc{F}_{\mu\nu}=\mathbb{L}_\Lambda F_{\mu\nu}.
\ee

The next step is to introduce the field strength for the 2-form gauge field $B_{\mu\nu}$, so that the kinetic term for $B_{\mu\nu}$ can be added into the action. We define that 3-form field strength $H_{\mu\nu\rho}$ by the following Bianchi identity:
\be
3\mc{D}_{[\mu}\mc{F}_{\nu\rho]}=\wtl H_{\mu\nu\rho}.
\ee
We expand
\be
\bsp
3\mc{D}_{[\mu}\mc{F}_{\nu\rho]}\ =\ &3\mc{D}_{[\mu}(2\ptl_\nu A_{\rho]}-\mathbb{L}_{A_\nu}A_{\rho]}+\wtl B_{\nu\rho]})\\[0.5ex]
\ =\ &-6\mathbb{L}_{A_{[\mu}}\ptl_\nu A_{\rho]}-3\ptl_{[\mu}(\mathbb{L}_{A_\nu}A_{\rho]})+3\mathbb{L}_{A_{[\mu}}\mathbb{L}_{A_\nu}A_{\rho]}+3\wtl\mc{D}_{[\mu}B_{\nu\rho]}\\[0.5ex]
\ =\ &-3\mathbb{L}_{A_{[\mu}}\ptl_\nu A_{\rho]}-3\mathbb{L}_{\ptl_{[\nu}A_\rho}A_{\mu]}+\wtl(A_{[\mu}\bullet[A_\nu,A_{\rho]}]_E)+3\wtl\mc{D}_{[\mu}B_{\nu\rho]}\\[0.5ex]
\ =\ &\wtl(3\mc{D}_{[\mu}B_{\nu\rho]}-3A_{[\mu}\bullet\ptl_\nu A_{\rho]}+A_{[\mu}\bullet[A_\nu,A_{\rho]}]_E).
\end{split}
\ee
In the first line we rewrote the E-bracket as generalized Lie derivative using (\ref{antisym}). From the first line to second line we used the fact that $\wtl$ commutes with $\mc{D}_\mu$. From the second line to third line we used the Jacobi identity (\ref{Jacobi2}). Finally from the third line to the fourth line we used (\ref{AAprop}). From this we can write out the form of ``naive'' 3-form field strength:
\be
H_{\mu\nu\rho}=3\mc{D}_{[\mu}B_{\nu\rho]}-3A_{[\mu}\bullet\ptl_\nu A_{\rho]}+A_{[\mu}\bullet[A_\nu,A_{\rho]}]_E
\ee

This field strength $H_{\mu\nu\rho}$ is also not gauge covariant. The way to fix it is to add a 3-form gauge field $C_{\mu\nu\rho}\in\mathfrak{C}(3\omega)$, and introduce the modified $\mc{H}_{\mu\nu\rho}$:
\be
\mc{H}_{\mu\nu\rho}=H_{\mu\nu\rho}+\wtl C_{\mu\nu\rho}.
\ee
Since $\wtl^2=0$, this $\mc{H}_{\mu\nu\rho}$ still satisfies the Bianchi identity
\be
3\mc{D}_{[\mu}\mc{F}_{\nu\rho]}=\wtl \mc{H}_{\mu\nu\rho}.
\ee
The general variation of $H_{\mu\nu\rho}$ can be collected in the following suggestive form:
\be
\bsp
\delta {\mc H}_{\mu\nu\rho} \ = \ &3\mc{D}_{[\mu}\delta B_{\nu\rho]}-3\mathbb{L}_{\delta A_{[\mu}}B_{\nu\rho]}-
3\delta A_{[\mu}\bullet \ptl_\nu A_{\rho]}-3A_{[\mu}\bullet \ptl_\nu\delta A_{\rho]}\\[0.5ex]
&+\delta A_{[\mu}\bullet[A_\nu,A_{\rho]}]_{E}
+2A_{[\mu}\bullet[\delta A_\nu,A_{\rho]}]_{E}+\wtl\delta C_{\mu\nu\rho}\\[0.5ex]
\ =\ &
3\mc{D}_{[\mu}\Delta B_{\nu\rho]}-3\delta A_{[\mu}\bullet\mc{F}_{\nu\rho]}+\wtl\delta C_{\mu\nu\rho}
-3\widehat\partial(\delta A_{[\mu}\bullet B_{\nu\rho]}) \\[0.5ex]
&+\delta A_{[\mu}\bullet\mathbb{L}_{A_\nu}A_{\rho]}+A_{[\mu}\bullet\mathbb{L}_{\delta A_\nu}A_{\rho]}-2A_{[\mu}\bullet\mathbb{L}_{A_\nu}\delta A_{\rho]}.
\end{split}
\ee
In the above derivation we made use of the identities mentioned before. Now we rewrite the terms in the last line above in a $\wtl$-exact form, using the following lemma: for any $A_\mu,A_\nu,C\in\mathfrak{A}(\omega)$:
\be
\bsp
\mathbb{L}_{A_{[\mu}}A_{\nu]}\bullet C+A_{[\mu}\bullet\mathbb{L}_C A_{\nu]}-2A_{[\mu}\bullet\mathbb{L}_{A_{\nu]}}C \ &= \ \mathbb{L}_{A_{[\mu}}(A_{\nu]}\bullet C)-A_{[\mu}\bullet\mathbb{L}_C A_{\nu]}-A_{[\mu}\bullet\mathbb{L}_{A_{\nu]}}C\\[0.5ex]
\ &= \ \mathbb{L}_{A_{[\mu}}(A_{\nu]}\bullet C)-A_{[\mu}\bullet\widehat\partial(A_{\nu]}\bullet C)\\[0.5ex]
\ &= \ \widehat\partial(A_{[\mu}\bullet(A_{\nu]}\bullet C)) .
\label{3identity}
\end{split}
\ee
In the last step we made use of the magic formula (\ref{magic}), with $\Lambda=A_\mu$, $X=A_\nu\bullet C$. Then we insert $C=\delta A_\rho$, and get the final answer for $\delta \mc{H}_{\mu\nu\rho}$:
\be
\delta\mc{H}_{\mu\nu\rho}=3\mc{D}_{[\mu}\Delta B_{\nu\rho]}-3\delta A_{[\mu}\bullet\mc{F}_{\nu\rho]}+\wtl\Delta C_{\mu\nu\rho},
\ee
where we have defined
\be
\Delta C_{\mu\nu\rho}\equiv\delta C_{\mu\nu\rho}-3\delta A_{[\mu}\bullet B_{\nu\rho]}+A_{[\mu}\bullet(A_\nu\bullet \delta A_{\rho]}).
\ee
Specify to the gauge variation $\delta_\Lambda$, with the rules $\delta_\Lambda A_\mu=\mc{D}_\mu\Lambda$, $\Delta_\Lambda B_{\mu\nu}=\Lambda\bullet\mc{F}_{\mu\nu}$, the failure of gauge covariance is
\be
\bsp
\delta_\Lambda\mc{H}_{\mu\nu\rho}-\mathbb{L}_{\Lambda}\mc{H}_{\mu\nu\rho}\ =\ &3\mc{D}_{[\mu}(\Lambda\bullet\mc{F}_{\nu\rho]})-3\mc{D}_{[\mu}\Lambda\bullet\mc{F}_{\nu\rho]}+\wtl\Delta C_{\mu\nu\rho}-\mathbb{L}_{\Lambda}\mc{H}_{\mu\nu\rho}\\[0.5ex]
\ =\ &3\Lambda\bullet\mc{D}_{[\mu}\mc{F}_{\nu\rho]}+\wtl\Delta C_{\mu\nu\rho}-\mathbb{L}_{\Lambda}\mc{H}_{\mu\nu\rho}\\[0.5ex]
\ =\ &\Lambda\bullet\wtl\mc{H}_{\mu\nu\rho}-\mathbb{L}_{\Lambda}\mc{H}_{\mu\nu\rho}+\wtl\Delta C_{\mu\nu\rho}\\[0.5ex]
\ =\ &\wtl(\Delta C_{\mu\nu\rho}-\Lambda\bullet\mc{H}_{\mu\nu\rho}).
\end{split}
\ee
In the last step we used the magic formula (\ref{magic}) again, with $X=\mc{H}_{\mu\nu\rho}$. Now if we assign
\be
\Delta C_{\mu\nu\rho}=\Lambda\bullet\mc{H}_{\mu\nu\rho},
\ee
then the field strength $\mc{H}_{\mu\nu\rho}$ is gauge covariant and can be directly implemented in the action. Note that indeed the identities grouped as (\ref{Forms2}) and (\ref{Forms3}) are repetitively used. 

The tensor hierarchy will continue, and more identities, (\ref{Forms4})(\ref{Forms5}), will naturally play a role in the derivation. We will not present the full detail, but we collect all the essential formulas here.

The gauge fields $A_\mu$, $B_{\mu\nu}$, $C_{\mu\nu\rho}$, $D_{\mu\nu\rho\sigma}$, $E_{\mu\nu\rho\sigma\tau}$ are elements of $\mathfrak{A}(\omega)$, $\mathfrak{B}(2\omega)$, $\mathfrak{C}(3\omega)$, $\mathfrak{D}(4\omega)$, $\mathfrak{E}(5\omega)$. Their corresponding covariant field strengths are in the same representation of $G$, and we list their form below (up to 5-form field strength):
\be
\bsp
\mc{F}_{\mu\nu}\ =\ &2\ptl_{[\mu}A_{\nu]}-[A_\mu,A_\nu]_E+\wtl B_{\mu\nu},\\[0.5ex]
\mc{H}_{\mu\nu\rho}\ =\ &3\mc{D}_{[\mu} B_{\nu\rho]}-3\ptl_{[\mu}A_\nu\bullet A_{\rho]}+A_{[\mu}\bullet[A_\nu,A_{\rho]}]_E+\wtl C_{\mu\nu\rho},\\[0.5ex]
\mc{J}_{\mu\nu\rho\sigma}\ =\ &4D_{[\mu}C_{\nu\rho\sigma]}+3\wtl B_{[\mu\nu}\bullet B_{\rho\sigma]}-6\mc{F}_{[\mu\nu}\bullet B_{\rho\sigma]}+4A_{[\mu}\bullet(A_\nu\bullet\ptl_\rho A_{\sigma]})\\[0.5ex]
&-A_{[\mu}\bullet(A_\nu\bullet[A_\rho,A_{\sigma]}]_E)+\wtl D_{\mu\nu\rho\sigma},\\[0.5ex]
K_{\mu\nu\rho\sigma\tau} \ = \ \, &5\,\mc{D}_{[\mu}D_{\nu\rho\sigma\tau]}+15\, B_{[\mu\nu}\bullet\mc{D}_\rho B_{\sigma\tau]}-10\, \mc{F}_{[\mu\nu}\bullet C_{\rho\sigma\tau]}\\[0.5ex]
&+30\,B_{[\mu\nu}\bullet(-A_\rho\bullet\ptl_\sigma A_{\tau]}+\fracs{1}{3}A_\rho\bullet[A_\sigma,A_{\tau]}]_E)\\[0.5ex]
&-5\,A_{[\mu}\bullet (A_\nu\bullet (A_\rho\bullet\ptl_\sigma A_{\tau]}))+A_{[\mu}\bullet (A_\nu\bullet (A_\rho\bullet[A_\sigma,A_{\tau]}]_E))+\wtl E_{\mu\nu\rho\sigma\tau}.\label{fstrengths}
\end{split}
\ee
They satisfy the following Bianchi identities:
\be
3\mc{D}_{[\mu}\mc{F}_{\nu\rho]}=\wtl \mc{H}_{\mu\nu\rho},\label{Bianchi3}
\ee
\be
4\mc{D}_{[\mu}\mc{H}_{\nu\rho\sigma]}+3\mc{F}_{[\mu\nu}\bullet\mc{F}_{\rho\sigma]}=\wtl \mc{J}_{\mu\nu\rho\sigma},\label{Bianchi4}
\ee
\be
5\mc{D}_{[\mu}\mc{J}_{\nu\rho\sigma\tau]}+10\mc{F}_{[\mu\nu}\bullet\mc{H}_{\rho\sigma\tau]}=\wtl \mc{K}_{\mu\nu\rho\sigma\tau}.\label{Bianchi5}
\ee

The variation of the field strengths can be simplified by introducing the following ``covariant variations'':
\be
\bsp
\Delta B_{\mu\nu}\ =\ &\delta B_{\mu\nu}+A_{[\mu}\bullet A_{\nu]},\\[0.5ex]
\Delta C_{\mu\nu\rho}\ =\ &\delta C_{\mu\nu\rho}-3\delta A_{[\mu}\bullet B_{\nu\rho]}+A_{[\mu}\bullet (A_\nu\bullet \delta A_{\rho]}),\\[0.5ex]
\Delta D_{\mu\nu\rho\sigma}\ =\ &\delta D_{\mu\nu\rho\sigma}-4\delta A_{[\mu}\bullet C_{\nu\rho\sigma]}+3B_{[\mu\nu}\bullet(\delta B_{\rho\sigma]}+2A_\rho\bullet\delta A_{\sigma]})+A_{[\mu}\bullet(A_\nu\bullet(A_\rho\bullet\delta A_{\sigma]})),\\[0.5ex]
 \Delta E_{\mu\nu\rho\sigma\tau} \ = \ \, &\delta E_{\mu\nu\rho\sigma\tau}-5\,\delta A_{[\mu}\bullet 
  D_{\nu\rho\sigma\tau]}-10\,\delta B_{[\mu\nu}\bullet C_{\rho\sigma\tau]}\\[0.5ex] 
  & -15\, B_{[\mu\nu}\bullet(\delta A_\rho\bullet B_{\sigma\tau]})
  -10\, (A_{[\mu}\bullet \delta A_\nu)\bullet C_{\rho\sigma\tau]}\\[0.5ex]
 & 
 +10\, B_{[\mu\nu}\bullet (A_\rho\bullet(A_\sigma\bullet\delta A_{\tau]}))+A_{[\mu}\bullet (A_\nu\bullet (A_\rho\bullet (A_\sigma\bullet \delta A_{\tau]}))).\label{covv}
\end{split}
\ee
The general variation of covariant field strengths, written in terms of the covariant variations are:
\be
\bsp
\delta \mc{F}_{\mu\nu}\ =\ &2\mc{D}_{[\mu}\delta A_{\nu]}+\wtl\Delta B_{\mu\nu},\\[0.5ex]
\delta \mc{H}_{\mu\nu\rho}\ =\ &3\mc{D}_{[\mu}\Delta B_{\nu\rho] }-3\delta A_{[\mu}\bullet\mc{F}_{\nu\rho]}+\wtl \Delta C_{\mu\nu\rho},\\[0.5ex]
\delta\mc{J}_{\mu\nu\rho\sigma}\ =\ &4\mc{D}_{[\mu}\Delta C_{\nu\rho\sigma]}^\delta-4\delta A_{[\mu}\bullet\mc{H}_{\nu\rho\sigma]}-6\mc{F}_{[\mu\nu}\bullet\Delta B_{\rho\sigma] }+\wtl\Delta D_{\mu\nu\rho\sigma},\\[0.5ex]
\delta {\cal K}_{\mu\nu\rho\sigma\tau} \ = \ \, &
  5\, {\cal D}_{[\mu}\Delta D_{\nu\rho\sigma\tau]}-5\,\delta A_{[\mu}\bullet {\cal J}_{\nu\rho\sigma\tau]}
  -10\,{\cal F}_{[\mu\nu}\bullet \Delta C_{\rho\sigma\tau]},\\[0.5ex]
  &-10\,{\cal H}_{[\mu\nu\rho}\bullet \Delta B_{\sigma\tau]}
  +\widehat\partial(\Delta E_{\mu\nu\rho\sigma\tau}).\label{varifields}
\end{split}
\ee

There are a set of gauge transformations associated to different gauge fields. We denote them by $\delta_\Lambda,\delta_\Xi,\delta_\Theta,\delta_\Omega$. The gauge parameters are $\Lambda\in\mathfrak{A}(\omega)$, $\Xi\in\mathfrak{B}(2\omega)$, $\Theta\in\mathfrak{C}(3\omega)$, $\Omega\in\mathfrak{D}(4\omega)$, $\Upsilon\in\mathfrak{E}(5\omega)$. The transformation rules on the gauge fields, in terms of covariant variations are:
\be
  \begin{split}
    \delta A_\mu \ &= \ \mathcal{D}_\mu \Lambda - \widehat\partial\, \Xi_{\mu}\;, \\[0.5ex]
    \Delta B_{\mu\nu} \ &= \  2\,\mc{D}_{[\mu}\Xi_{\nu]}
    +\Lambda\bullet {\cal F}_{\mu\nu}-\widehat\partial\Theta_{\mu\nu}\;, \\[0.5ex]
    \Delta C_{\mu\nu\rho} \ &= \ 3\,{\cal D}_{[\mu}\Theta_{\nu\rho]}
    +\Lambda\bullet {\cal H}_{\mu\nu\rho}
    + 3\,{\cal F}_{[\mu\nu}\bullet \Xi_{\rho]}
    -\widehat\partial\Omega_{\mu\nu\rho} \\[0.5ex]
    \Delta D_{\mu\nu\rho\sigma} \ &= \ 4\,{\cal D}_{[\mu}\Omega_{\nu\rho\sigma]}
    +\Lambda \bullet \mc{J}_{\mu\nu\rho\sigma}
    - 4\, \mc{H}_{[\mu\nu\rho}\bullet\,\Xi_{\sigma]}  + 6\,\mc{F}_{[\mu\nu}\bullet\,\Theta_{\rho\sigma]}
    - \widehat\partial\Upsilon_{\mu\nu\rho\sigma}\;, \\[0.5ex]
    \Delta E_{\mu\nu\rho\sigma\tau} \ &= \ 5\,\mc{D}_{[\mu}\Upsilon_{\nu\rho\sigma\tau]}
    +\Lambda\bullet {\cal K}_{\mu\nu\rho\sigma\tau}-5\,{\cal J}_{[\mu\nu\rho\sigma}\bullet \Xi_{\tau]} \\[0.5ex]
  &\qquad -10\,{\cal H}_{[\mu\nu\rho}\bullet \Theta_{\sigma\tau]}+10\,{\cal F}_{[\mu\nu}\bullet \Omega_{\rho\sigma\tau]}
  +\cdots \;.
   \end{split}
  \ee

\section{SL$(2,\mathbb{R})\times \mathbb{R}_+$}

Now we study the specific groups. The highest dimension where an EFT may exist is 9, with symmetry group $G=$SL$(2,\mathbb{R})\times \mathbb{R}_+$. The representation of 1-form gauge field corresponds to the interior directions of the EFT, given in Table \ref{t:reps}, which suggests that the interior dimension of $d=9$ EFT is 3. We decompose the 3 interior dimensions $M=1,2,3$ into $\alpha,\beta,\dots=1,2$, which label the SL(2) doublet, and $z=3$, which labels the SL(2) singlet. The interior coordinates are denoted by $Y^M$.

The only primary invariant tensor of $G=$SL$(2,\mathbb{R})\times \mathbb{R}_+$ is the Kronecker delta tensor $\delta^\alpha_\beta$, $\delta^z_z$. Hence in the construction we do not use tensors other than this.

Now we want to impose a section condition. As usual, the solution to the section condition should contain the $9+2$ dimensional M-theory solution and the $9+1$ dimensional IIB solution. With this observation we propose the following section condition:
\be
\ptl_\alpha\otimes\ptl_z=0\ (\alpha=1,2).\label{section9}
\ee
It is easy to observe that the solution $\ptl_z=0$, $\ptl_\alpha\neq0$ corresponds to the M-theory solution, and the solution  $\ptl_z\neq 0$, $\ptl_\alpha=0$ corresponds to the IIB solution. This construction actually resembles the early postulation of M/II-B duality\cite{Schwarz:1995jq}\cite{Schwarz:1996bh}.

The only non-vanishing components of $Z$-tensor are:
\be
Z^{\alpha z}{}_{\beta z}=Z^{\alpha z}{}_{z\beta}=Z^{z\alpha}{}_{\beta z}=Z^{z\alpha}{}_{z\beta}=\delta^\alpha_\beta.
\ee

With this choice of $Z$-tensor and formula (\ref{gdvomega}), the generalized Lie derivative acting on the vector with specific weight $\omega$ is\footnote{$\omega$ will be determined below, after (\ref{LAF})}
\be
\bsp
\mathbb{L}_\Lambda V^\alpha\ =\ &\Lambda^\beta\ptl_\beta V^\alpha+\Lambda^z\ptl_z V^\alpha-V^\beta\ptl_\beta\Lambda^\alpha+\ptl_z\Lambda^z V^\alpha\\[0.5ex]
\mathbb{L}_\Lambda V^z\ =\ &\Lambda^\beta\ptl_\beta V^z+\Lambda^z\ptl_z V^z-V^z\ptl_z\Lambda^z+\ptl_\alpha\Lambda^\alpha V^z.\label{gdp9}
\end{split}
\ee

The E-bracket is given by: 
\be
\bsp
[U,V]_E^\alpha\ =\ &U^M\ptl_M V^\alpha+\fracs{1}{2}\ptl_z U^\alpha V^z+\fracs{1}{2}\ptl_z U^z V^\alpha-(U\leftrightarrow V),\\[0.5ex]
[U,V]_E^z\ =\ &U^M\ptl_M V^z+\fracs{1}{2}\ptl_\alpha U^z V^\alpha+\fracs{1}{2}\ptl_\alpha U^\alpha V^z-(U\leftrightarrow V).\label{Ebracket9}
\end{split}
\ee

One can check that with these definitions, along with the section condition (\ref{section9}), the gauge closure condition indeed holds. For the higher tensor object, a hint is to look at the corresponding representation in 9D maximal supergravity, e.g. Table 2.4 of \cite{Roest}. The next tensor object in $\mathfrak{B}(2\omega)$ is a doublet, it is formally written as:
\be
B^{\alpha,z}\in\mathfrak{B}(2\omega).
\ee
The generalized Lie derivative on this object can be directly constructed by applying (\ref{gdp9}) to a tensor with two indices. We write it down explicitly:
\be
\bsp
\mathbb{L}_\Lambda B^{\alpha,z}\ =\ &\Lambda^M\ptl_M B^{\alpha,z}-B^{\beta,z}\ptl_\beta\Lambda^\alpha-B^{\alpha,z}\ptl_z\Lambda^z+\ptl_M\Lambda^M B^{\alpha,z}\\[0.5ex]
\ =\ &\Lambda^M\ptl_M B^{\alpha,z}-B^{\beta,z}\ptl_\beta\Lambda^\alpha+B^{\alpha,z}\ptl_\beta\Lambda^\beta.
\end{split}
\ee

The $\bullet$ operator on $A_1,A_2\in\mathfrak{A}$ is defined to be:
\be
(A_1\bullet A_2)^{\alpha, z}=A_1^\alpha A_2^z+A_2^\alpha A_1^z.
\ee
The $\widehat\ptl $ operator on $B\in\mathfrak{B}(2\omega)$ is defined to be
\be
(\widehat\ptl  B)^\alpha=\ptl_z B^{\alpha,z}\ ,\ (\widehat\ptl  B)^z=\ptl_\alpha B^{\alpha,z}.
\ee
Since $B$ is already written in two-index form it is unnecessary to introduce the isomorphism $I_B$. We can directly verify that $\wtl B$ is a trivial gauge parameter, using the section condition (\ref{section9}):
\be
\bsp
\mathbb{L}_{\wtl B} A^\alpha\ =\ &\ptl_z B^{\beta,z}\ptl_\beta A^\alpha+\ptl_\beta B^{\beta,z}\ptl_z A^\alpha-A^\beta\ptl_\beta\ptl_z B^{\alpha,z}+\ptl_z\ptl_\beta B^{\beta,z}A^\alpha\\
\ =\ &0\\
\mathbb{L}_{\wtl B} A^z\ =\ &\ptl_z B^{\beta,z}\ptl_\beta A^z+\ptl_\alpha B^{\alpha,z}\ptl_z A^z-A^z\ptl_z\ptl_\beta B^{\beta,z}+\ptl_\alpha\ptl_z B^{\alpha,z}A^z\\
\ =\ &0.
\end{split}
\ee

The tensors in $\mathfrak{C}(3\omega)$ are singlets. We denote them by $C^{[\alpha\beta]z}$. This notation is a bit redundant, but it helps to simplify the pool of generalized Lie derivative rules. We do not need to define any new rules apart from (\ref{gdp9}), which can be naturally extended to tensors with multiple indices. The $\bullet$ operator on $A\in\mathfrak{A}$, $B\in\mathfrak{B}$ is defined as:
\be
(A\bullet B)^{[\alpha\beta]z}=(B\bullet A)^{[\alpha\beta]z}=2A^{[\alpha}B^{\beta]z}.
\ee
The $\widehat\ptl $ operator on $C\in\mathfrak{C}(3\omega)$ is defined to be
\be
(\widehat\ptl  C)^{\alpha,z}=\ptl_\beta C^{[\beta\alpha]z}.
\ee

The next tensor set $\mathfrak{D}(4\omega)$ also represents singlets. The elements are denoted by $D^{[\alpha\beta]zz}$. The bullet operators on $A\in\mathfrak{A}$, $B_1,B_2\in\mathfrak{B}$, $C\in\mathfrak{C}$ are:
\be
(A\bullet C)^{[\alpha\beta]zz}=(C\bullet A)^{[\alpha\beta]zz}=A^z C^{[\alpha\beta]z},
\ee
\be
(B_1\bullet B_2)^{[\alpha\beta]zz}=2B_1^{[\alpha|z}B_2^{|\beta]z}.
\ee
The $\widehat\ptl $ operator on $D\in\mathfrak{D}(4\omega)$ is
\be
(\widehat\ptl  D)^{[\alpha\beta]z}=\ptl_z D^{[\alpha\beta]zz}.
\ee

Note that although $\mathfrak{C}(3\omega)$ and $\mathfrak{D}(4\omega)$ are singlets, they do not transform as scalars under $\mathbb{L}_\Lambda$. Applying the generalized Lie derivative (\ref{gdp9}) to tensors with multiple indices, we obtain:
\be
\mathbb{L}_\Lambda C^{[\alpha\beta]z}=\Lambda^M\ptl_M C^{[\alpha\beta]z}+\ptl_z\Lambda^z C^{[\alpha\beta]z},
\ee
\be
\mathbb{L}_\Lambda D^{[\alpha\beta]zz}=\Lambda^M\ptl_M D^{[\alpha\beta]zz}+\ptl_\gamma\Lambda^\gamma D^{[\alpha\beta]zz},
\ee
which are not the rules for scalars. But if one solves the section condition explicitly, then for the M-theory solution where $\ptl_z=0$, $C^{[\alpha\beta]z}$ is a scalar. For the IIB solution where $\ptl_\alpha=0$, $D^{[\alpha\beta]zz}$ is a scalar.

The next tensor $E\in\mathfrak{E}(5\omega)$ is a doublet. It is denoted by $E^{\gamma[\alpha\beta]zz}$. The bullet operators on $A\in\mathfrak{A}$, $B\in\mathfrak{B}$, $C\in\mathfrak{C}$, $D\in\mathfrak{D}$ are:
\be
(A\bullet D)^{\gamma[\alpha\beta]zz}=(D\bullet A)^{\gamma[\alpha\beta]zz}=A^\gamma D^{[\alpha\beta]zz},
\ee
\be
(B\bullet C)^{\gamma[\alpha\beta]zz}=(C\bullet B)^{\gamma[\alpha\beta]zz}=B^{\gamma,z}C^{[\alpha\beta]z}.
\ee
The $\widehat\ptl $ operator on $E\in\mathfrak{E}(5\omega)$ is
\be
(\widehat\ptl  E)^{[\alpha\beta]zz}=\ptl_\gamma E^{\gamma[\alpha\beta]zz}.
\ee

Finally, $F\in\mathfrak{F}(6\omega)$ is in the same representation as $\mathfrak{A}(\omega)$, one can write its components as
\be
F^z\equiv F^{[\delta\gamma][\alpha\beta]zz}\ ,\ F^\gamma\equiv F^{z\gamma[\alpha\beta]zz}.
\ee
The relevant rules of $\bullet$ operators and $\wtl$ can be defined as:
\be
\bsp
&(A\bullet E)^z=(E\bullet A)^z=A^{[\delta}E^{\gamma][\alpha\beta]zz},\\
&(A\bullet E)^\gamma=(E\bullet A)^\gamma=A^{z}E^{\gamma[\alpha\beta]zz},\\
&(B\bullet D)^\gamma=(D\bullet B)^\gamma=B^{\gamma,z}D^{[\alpha\beta]zz},\\
&(B\bullet D)^z=(D\bullet B)^z=0,\\
&(C_1\bullet C_2)^z=C_1^{[\alpha\beta]z}C_2^{[\gamma\delta]z},\\
&(C_1\bullet C_2)^\gamma=0,\\
&(\wtl F)^{\gamma[\alpha\beta]zz}=\ptl_z F^{z\gamma[\alpha\beta]zz}=\ptl_z F^\gamma.
\end{split}
\ee

We can explicitly check that $A^z F^z$ and $A^{[\alpha} F^{\beta]}$ are scalars with weight 1. To do this, we first simplify the form of generalized Lie derivative acting on $F^M$:
\be
\bsp
\mathbb{L}_\Lambda F^z\ =\ &\Lambda^M\ptl_M F^z-\ptl_\tau\Lambda^{[\delta|}F^{\tau|\gamma][\alpha\beta]zz}-\ptl_\tau\Lambda^{[\gamma}F^{\delta]\tau[\alpha\beta]zz}-\ptl_\tau\Lambda^{[\alpha|}F^{[\delta\gamma]\tau|\beta]zz}-\ptl_\tau\Lambda^{[\beta|}F^{[\delta\gamma]|\alpha]\tau zz}\\[0.5ex]
&+4\ptl_z\Lambda^z F^z+2(\ptl_\tau\Lambda^\tau F^z-\ptl_z\Lambda^z F^z)\\[0.5ex]
\ =\ &\Lambda^M\ptl_M F^z+2\ptl_z\Lambda^z F^z.
\end{split}
\ee
Hence
\be
\mathbb{L}_\Lambda (A^z F^z)=A^z\mathbb{L}_\Lambda F^z+F^z\mathbb{L}_\Lambda A^z=A^M\ptl_M(A^z F^z) +\ptl_M\Lambda^M A^z F^z.
\ee
Similarly
\be
\mathbb{L}_\Lambda F^\alpha=\Lambda^M\ptl_M F^\alpha+F^\alpha\ptl_z\Lambda^z+2F^{\alpha}\ptl_\beta\Lambda^{\beta}-F^\beta\ptl_\beta\Lambda^\alpha,
\ee
and then
\be
\mathbb{L}_\Lambda (A^{[\alpha} F^{\beta]})=A^{[\alpha}\mathbb{L}_\Lambda F^{\beta]}+\mathbb{L}_\Lambda A^{[\alpha} F^{\beta]}=\Lambda^M\ptl_M(A^{[\alpha} F^{\beta]}) +\ptl_M\Lambda^M A^{[\alpha} F^{\beta]}.\label{LAF}
\ee
Similarly, it is not hard to check that $C^{[\alpha\beta]z}D^{[\gamma\delta]zz}$ and $B^{[\alpha|,z}E^{|\beta][\gamma\delta]zz}$ are also scalars with weight 1.

From this observation, we obtain that $\omega=1/7=1/(d-2)$, which is consistent with the general observation in section 2. The length of exceptional exact sequence is $l=6=d-3$.

One can explicitly check that all the identities in section 2 hold. For example, we want to prove the magic formula (\ref{magic}) for $X=C\in\mathfrak{C}(3\omega)$. We write out 
\be
\bsp
\mathbb{L}_\Lambda C^{[\alpha\beta]z}-(\Lambda\bullet\wtl C)^{[\alpha\beta]z}\ =\ &\Lambda^\gamma\ptl_\gamma C^{[\alpha\beta]z}-C^{[\gamma\beta]z}\ptl_\gamma\Lambda^\alpha-C^{[\alpha\gamma]z}\ptl_\gamma\Lambda^\beta+\ptl_z \Lambda^z C^{[\alpha\beta]z}\\[0.5ex]
&+\ptl_\gamma\Lambda^\gamma C^{[\alpha\beta]z}
-2\Lambda^{[\alpha}\ptl_\gamma C^{[\gamma\beta]]z}\\[0.5ex]
\ =\ &\ptl_z(\Lambda^z C^{[\alpha\beta]z})\ =\ \wtl(\Lambda\bullet C)^{[\alpha\beta]z}.
\end{split}
\ee

Another simple example is the identity (\ref{AAAprop}), which is equivalent to
\be
\bsp
&A_1^{[\alpha}A_2^{(\beta]}A_3^{z)}+A_2^{[\alpha}A_3^{(\beta]}A_1^{z)}+A_3^{[\alpha}A_1^{(\beta]}A_2^{z)}
\ =\ A_1^{([\alpha}A_2^{\beta]}A_3^{z)}
\ =\ 0.
\end{split}
\ee

Finally we make some comments on the Poincar\'{e} lemma in $d=9$ EFT. If for $B^\alpha\equiv B^{\alpha,z}\in\mathfrak{B}(2\omega)$, $\wtl B=0$, that is, \be
\ptl_\alpha B^{\alpha}=\ptl_z B^{\alpha}=0,
\ee
can we conclude locally $B^\alpha=\ptl_\beta C^{[\beta\alpha]z}$, where $C^{[\beta\alpha]z}\in\mathfrak{C}(3\omega)$?

 In the M-theory solution we already know $\ptl_z=0$, then one simply applies the usual Poincar\'{e} lemma in two-dimensional space $\{Y^1,Y^2\}$. However, in the IIB solution $\ptl_\alpha=0$, the logic breaks down, since apparently $B^\alpha$ can be a non-vanishing constant value, however, $\wtl C= \ptl_\beta C^{[\beta\alpha]z}$ always vanishes, so the Poincar\'{e} lemma does not follow. 
 
 Then we analyze if for $C^{[\alpha\beta]z}\in\mathfrak{C}(3\omega)$, $\wtl C=0$, can we conclude that locally $C=\wtl D$. For the M-theory solution $\ptl_z=0$, $\wtl D$ always vanishes, hence the Poincar\'{e} lemma cannot hold. But this works for the IIB solution, since $\wtl C$ always vanish and one can always locally write $C=\ptl_z D$.
 
 We summarize the validity of Poincar\'{e} lemma in Table \ref{t:Poincare9}.
 
 It is notable that only half of the Poincar\'{e} lemmas work, in an alternating pattern. 
 
 \begin{table}
 \centering
 \begin{tabular}{|c|c|c|c|}
 \hline
  & M-theory solution & IIB solution\\
  \hline
 $\wtl B=0\rightarrow B=\wtl C$?  & Yes & No\\
 $\wtl C=0\rightarrow C=\wtl D$?  & No & Yes\\
 $\wtl D=0\rightarrow D=\wtl E$?& Yes & No\\
 $\wtl E=0\rightarrow E=\wtl F$?  & No & Yes\\
\hline
\end{tabular}
\caption{Validity of Poincar\'{e} lemma in $d=9$ EFT, here $B\in\mathfrak{B}(2\omega)$, $C\in\mathfrak{C}(3\omega)$, $D\in\mathfrak{D}(4\omega)$, $E\in\mathfrak{E}(5\omega)$, $F\in\mathfrak{F}(6\omega)$, as usual.}\label{t:Poincare9}
\end{table}

The notion of Poincar\'{e} lemma here is not directly related to the statement in usual differential geometry, that the contractible manifold has trivial cohomology. To relate these two version of Poincar\'{e} lemma, one needs a Stokes' theorem in this new differential geometry. However, the naive replacement of the exterior derivative by $\wtl$ does not work. For this group SL$(2,\mathbb{R})\times\mathbb{R}_+$ the length of exceptional chain complex is even greater than the dimension of interior manifold. The notion of higher form cannot be interpreted in any conventional way.

\section{SL$(3,\mathbb{R})\times$SL$(2,\mathbb{R})$}

We review the generalized Cartan calculus in 8 dimensions\cite{8D} for completeness. 

The group $G=$SL$(3,\mathbb{R})\times$SL$(2,\mathbb{R})$, and the vector and interior directions are in $(3,2)$ representation, denoted by $i\alpha$, where $i=1,2,3$ and $\alpha=1,2$ labels SL(3) and SL(2) fundamental indices respectively. The length of exceptional chain complex is 5, and the specific weight $\omega=1/6$. 

The $Z$-tensor is
\be
Z^{i\alpha,j\beta}{}_{k\gamma,l\delta}=\epsilon^{ijm}\epsilon_{klm}\epsilon^{\alpha\beta}\epsilon_{\gamma\delta}.
\ee
The section condition is then
\be
\epsilon^{ijm}\epsilon^{\alpha\beta}\ptl_{i\alpha}\otimes\ptl_{j\beta}=0.
\ee
The 8+3D M-theory solution is given by the choice $\ptl_{11},\ptl_{21},\ptl_{31}\neq 0$, $\ptl_{12}=\ptl_{22}=\ptl_{32}=0$, and the 8+2D IIB solution is given by $\ptl_{11},\ptl_{12}\neq 0$, others$=0$.

We list the tensor objects in Table \ref{t:tensor8}, with their representation in SL(3)$\times$SL(2) and weights.
 \begin{table}
 \centering
 \begin{tabular}{|c|c|c|c|}
 \hline
 Tensor space&rep.&notation with index&weight\\
 \hline
$\mathfrak{A}$&$(\mathbf{3},\mathbf{2})$& $A^{i\alpha}$ & $1/6$\\
$\mathfrak{B}$&$(\mathbf{\bar{3}},\mathbf{1})$& $B_i$ & $1/3$\\
$\mathfrak{C}$&$(\mathbf{1},\mathbf{2})$& $C^\alpha$ & $1/2$\\
$\mathfrak{D}$&$(\mathbf{3},\mathbf{1})$& $D^i$ & $2/3$\\
$\mathfrak{E}$&$(\mathbf{\bar{3}},\mathbf{2})$ & $E_{i\alpha}$ & $5/6$\\
\hline
\end{tabular}
\caption{The list of tensor objects in SL$(3,\mathbb{R})\times$SL$(2,\mathbb{R})$ EFT.}\label{t:tensor8}
\end{table}

We list the generalized Lie derivatives on these tensors: 
\bea
\mathbb{L}_\Lambda A^{i\alpha} &=& \Lambda^{j\beta}\ptl_{j\beta} A^{i\alpha}-A^{j\beta}\ptl_{j\beta}\Lambda^{i\alpha}+\epsilon^{ijm}\epsilon_{klm}\epsilon^{\alpha\beta}\epsilon_{\gamma\delta}\ptl_{j\beta}\Lambda^{k\gamma} A^{l\delta},\\[0.5ex]
 \mathbb{L}_\Lambda B_i &=& \Lambda^{j\alpha}\ptl_{j\alpha} B_i+\ptl_{i\alpha}\Lambda^{j\alpha}B_j,\\[0.5ex]
  \mathbb{L}_{\Lambda}C^{\alpha} &=& \Lambda^{i\beta}\partial_{i\beta}C^{\alpha}
   -\epsilon^{\alpha\beta}\epsilon_{\gamma\delta}\,\partial_{i\beta}\Lambda^{i\gamma}\,C^{\delta},\\[0.5ex]
   \mathbb{L}_{\Lambda} D^i &=& \Lambda^{j\gamma}\partial_{j\gamma}D^i
  -D^j\partial_{j\gamma}\Lambda^{i\gamma}+\partial_{j\gamma}\Lambda^{j\gamma}
  \,D^i\; ,\\[0.5ex]
  \mathbb{L}_\Lambda E_{i\alpha} &=& \Lambda^{j\beta}\ptl_{j\beta}E_{i\alpha}+E_{j\alpha}\ptl_{i\beta}\Lambda^{j\beta}
  +E_{i\beta}\ptl_{j\alpha}\Lambda^{j\beta}.
\eea
Then the operators $\bullet$ and $\wtl$ are defined as:
\bea
   (A_1\bullet A_2)_m&=&\epsilon_{ijm}\epsilon_{\alpha\beta}A_1^{i\alpha}A_2^{j\beta}\label{8AA},\\[0.5ex]
   (A\bullet B)^\alpha&=&B_{m}A^{m\alpha},\\[0.5ex]
   (A\bullet C)^m&=&\epsilon_{\alpha\beta}C^\alpha A^{m\beta},\\[0.5ex]
   (A\bullet D)_{m\alpha} &=&
   \epsilon_{mnk}\epsilon_{\alpha\beta}A^{n\beta}D^k,\\[0.5ex]
      (A\bullet E)&=&A^{i\alpha}E_{i\alpha},\\[0.5ex]
    (B_1\bullet B_2)^m &=&\epsilon^{ijm}B_{1i}B_{2j},\\[0.5ex]
   (B\bullet C)_{m\alpha} &=&
   \epsilon_{\alpha\beta}B_m C^\beta,\\[0.5ex]
   (B\bullet D)&=& B_m D^m,\\[0.5ex]
   (C_1\bullet C_2)&=&\epsilon_{\alpha\beta}C_1^\alpha C_2^\beta,\\[0.5ex]
   (\widehat\partial E)^m &=&\epsilon^{mnk}\epsilon^{\alpha\beta}\ptl_{n\alpha}E_{k\beta},\\[0.5ex]
   (\widehat\partial D)^\alpha &=&\epsilon^{\alpha\beta}\ptl_{m\beta}D^m,\\[0.5ex]
   (\widehat\partial C)_m&=&\ptl_{m\alpha}C^\alpha,\\[0.5ex]
   (\widehat\partial B)^{i\alpha}&=&\epsilon^{ijk}\epsilon^{\alpha\beta}\ptl_{j\beta}B_k.\label{8pB}
   \eea
As usual, the tensor objects labelled by $A,B,C,D,E$ are defined in tensor spaces $\mathfrak{A}(\omega)$, $\mathfrak{B}(2\omega)$, $\mathfrak{C}(3\omega)$, $\mathfrak{D}(4\omega)$, $\mathfrak{E}(5\omega)$ respectively.

The isomorphism $I_B$ that transforms $B^{i\alpha,j\beta}$ to $B_m$ is
\be
\bsp
I_B(B^{i\alpha,j\beta})_m\ =\ &\epsilon_{ijm}\epsilon_{\alpha\beta}B^{i\alpha,j\beta},\\[0.5ex]
I^{-1}_B(B_m)^{i\alpha,j\beta}\ =\ &\frac{1}{4}\epsilon^{ijm}\epsilon^{\alpha\beta}B_m.
\end{split}
\ee
These projectors explicitly lead to (\ref{8AA}) and (\ref{8pB}).

Finally we list the validity of Poincar\'{e} lemma (local exactness) in Table \ref{t:Poincare8}, in the same manner as the $d=9$ case.

 \begin{table}
 \centering
 \begin{tabular}{|c|c|c|c|}
 \hline
  & M-theory solution & IIB solution\\
  \hline
 $\wtl B=0\rightarrow B=\wtl C$? & Yes & No\\
 $\wtl C=0\rightarrow C=\wtl D$? & No & Yes\\
 $\wtl D=0\rightarrow D=\wtl E$? & Yes & No\\
\hline
\end{tabular}
\caption{Validity of Poincar\'{e} lemma in $d=8$ EFT, here $B\in\mathfrak{B}(2\omega)$, $C\in\mathfrak{C}(3\omega)$, $D\in\mathfrak{D}(4\omega)$, $E\in\mathfrak{E}(5\omega)$.}\label{t:Poincare8}
\end{table}

\section{SL$(5,\mathbb{R})$}

Maximally supersymmetric gravity in 7 dimensions has $G=$SL$(5,\mathbb{R})$ global symmetry. The 1-form gauge field is in rank-2 antisymmetric tensor representation of SL(5), hence the interior dimension of $d=7$ EFT is 10\cite{Berman:2010is}. We label them by $mn$, $m,n=1,\dots,5$ $(m\neq n)$. For any vector $V^{mn}$, $V^{mn}=-V^{nm}$ always holds. The form of generalized derivative which respects the gauge closure condition is given in \cite{Berman:2012vc}\cite{Musaev2013}\cite{MusaevThesis}:
\be
\bsp
\mathbb{L}_\Lambda V^{mn}\ =\ &\fracs{1}{2}\Lambda^{ij}\ptl_{ij}V^{mn}-\fracs{1}{2}V^{ij}\ptl_{ij}\Lambda^{mn}+\fracs{1}{8}\epsilon^{mnijv}\epsilon_{pqrsv}\ptl_{ij}\Lambda^{pq} V^{rs}\\[0.5ex]
\ =\ &\fracs{1}{2}(\Lambda^{ij}\ptl_{ij}V^{mn}+2\ptl_{ij}\Lambda^{im}V^{nj}+2\ptl_{ij}\Lambda^{nj}V^{im}+\ptl_{ij}\Lambda^{ij}V^{mn}).
\end{split}
\ee
Notice that there is an overall $\fracs{1}{2}$ factor, because there are only 10 interior dimensions, and each of them is counted twice in the summation.

The E-bracket easily follows:
\be
[U,V]_E^{mn}=\fracs{1}{2}U^{ij}\ptl_{ij}V^{mn}+\fracs{1}{16}\epsilon^{mnijv}\epsilon_{pqrsv}\ptl_{ij}U^{pq} V^{rs}-(U\leftrightarrow V).
\ee
The $Z$-tensor for $d=7$ EFT is:
\be
Z^{mn,ij}{}_{pq,rs}=\fracs{1}{4}\epsilon^{mnijv}\epsilon_{pqrsv},
\ee
and the section condition is given by
\be
\epsilon^{mnpqv}\ptl_{mn}\otimes\ptl_{pq}=0.
\ee
The two inequivalent maximal solutions to this section condition are:
\be
\ptl_{12},\ptl_{13},\ptl_{14},\ptl_{15}\neq 0,\ \mathrm{ others}=0
\ee
and
\be
\ptl_{12},\ptl_{13},\ptl_{23}\neq 0,\ \mathrm{ others}=0.
\ee
They corresponds to 7+4D M-theory solution and 7+3D IIB solution respectively\cite{Blair:2013gqa}.

The next tensor object is in (dual) vector representation of $SL(5)$: $B_v\in\mathfrak{B}(2\omega)$. The isomorphism from $B^{mn,pq}$ to $B_v$ is
\be
\bsp
(I_B B^{mn,pq})_v\ =\ &\fracs{1}{4}\epsilon_{mnpqv}B^{mn,pq}\\[0.5ex]
(I_B^{-1}B_v)^{mn,pq}\ =\ &\fracs{1}{6}\epsilon^{mnpqv}B_v.
\end{split}
\ee
The $\bullet$ operator results in such a tensor and the $\wtl$ acting on it are then defined as:
\be
 (A_1\bullet A_2)_v=\fracs{1}{4}\epsilon_{mnpqv}A_1^{mn}A_2^{pq},
 \ee
 \be
 (\wtl B)^{mn}=\fracs{1}{2}\epsilon^{mnpqv}\ptl_{pq}B_v.
 \ee
With these definitions, we can derive the rule of generalized Lie derivative acting on $B_v\in\mathfrak{B}(2\omega)$, using
\be
\bsp
\mathbb{L}_\Lambda (A_1\bullet A_2)_v\ =\ &(A_1\bullet \mathbb{L}_\Lambda A_2)_v+(1\leftrightarrow 2)\\[0.5ex]
\ =\ &\fracs{1}{8}\epsilon_{mnpqv}A_1^{mn}(\Lambda^{ij}\ptl_{ij}A_2^{pq}-\ptl_{ij}\Lambda^{pq}A_2^{ij}
+\fracs{1}{4}\epsilon^{pqijw}\epsilon_{rstuw}\ptl_{ij}\Lambda^{rs}A_2^{tu})+(1\leftrightarrow2)\\[0.5ex]
\ =\ &\fracs{1}{4}\Lambda^{ij}\ptl_{ij}(A_1\bullet A_2)_v+\fracs{1}{4}\ptl_{vm}\Lambda^{rs}A_1^{tu}A_2^{mn}\epsilon_{rstun}+(1\leftrightarrow2)\label{LAA7}
\end{split}
\ee
Then using the antisymmetrization property:
\be
\ptl_{v[m}\Lambda^{rs}A_1^{tu}A_2^{mn}\epsilon_{rstun]}=0,
\ee
one can rewrite (\ref{LAA7}) into
\be
\mathbb{L}_\Lambda (A_1\bullet A_2)_v=\fracs{1}{2}\Lambda^{ij}\ptl_{ij}(A_1\bullet A_2)_v+\fracs{1}{4}\epsilon_{mnrsp}A_1^{mn}A_2^{rs}\ptl_{vq}\Lambda^{pq}.
\ee
Hence we arrive at:
\be
(\mathbb{L}_\Lambda B)_v=\fracs{1}{2}\Lambda^{pq}\ptl_{pq}B_v+B_p\ptl_{vq}\Lambda^{pq}.
\ee

The next tensor object is vector $C^v\in\mathfrak{C}(3\omega)$. The relevant rules are:
\be
(A\bullet B)^v=(B\bullet A)^v=A^{vw}B_w,
\ee
\be
(\wtl C)_v=\ptl_{wv}C^w,
\ee
\be
(\mathbb{L}_\Lambda C)^v=\fracs{1}{2}\Lambda^{mn}\ptl_{mn}C^v-\ptl_{mn}\Lambda^{mv}C^n+\fracs{1}{2}\ptl_{mn}\Lambda^{mn}C^v.
\ee

Finally, for $D_{mn}\in\mathfrak{D}(4\omega)$, the relevant rules are:
\be
(A\bullet C)_{mn}=(C\bullet A)_{mn}=\fracs{1}{4}\epsilon_{mnpqv}A^{pq}C^v,
\ee
\be
(B_1\bullet B_2)_{mn}=B_{2[m}B_{1n]},
\ee
\be
(\wtl D)^v=\fracs{1}{2}\epsilon^{vmnpq}\ptl_{mn}D_{pq},
\ee
\be
\mathbb{L}_\Lambda D_{mn}=\fracs{1}{2}(\Lambda^{pq}\ptl_{pq}D_{mn}+D^{pq}\ptl_{pq}\Lambda_{mn}
-\fracs{1}{4}\epsilon_{mnpqv}\epsilon^{rstuv}\ptl_{rs}\Lambda^{pq}D_{tu}+\ptl_{pq}\Lambda^{pq}D_{mn}).
\ee
From the rules of generalized Lie derivative, it is clear that the scalars defined by index contraction: $A^{mn}D_{mn}$ and $B_v C^v$ are scalars with weight 1.

The length of exceptional chain complex is $l=4$, and the weight $\omega=1/5$. The additional identities listed at the end of  section 2 hold only when the $\bullet$ operations give tensor with weight smaller than 1.

The validity of Poincar\'{e} lemmas is listed in Table \ref{t:Poincare7}.

\begin{table}
 \centering
 \begin{tabular}{|c|c|c|c|}
 \hline
 & M-theory solution & IIB solution\\
  \hline
 $\wtl B=0\rightarrow B=\wtl C$?  & Yes & No\\
 $\wtl C=0\rightarrow C=\wtl D$? & No & Yes\\
\hline
\end{tabular}
\caption{Validity of Poincar\'{e} lemma in $d=7$ EFT, here $B\in\mathfrak{B}(2\omega)$, $C\in\mathfrak{C}(3\omega)$, $D\in\mathfrak{D}(4\omega)$.}\label{t:Poincare7}
\end{table}

\section{SO(5,5)}

For $d=6$ EFT, the symmetry group $G=$SO$(5,5)$, and the interior dimensions are labelled by the 16 components of SO$(5,5)$ Majorana-Weyl spinor: $\alpha,\beta,\dots=1,\dots,16$\footnote{Upon completion of this paper, the SO(5,5) EFT was constructed in\cite{6D}, which has some overlap with the discussions in this section}. The form of generalized Lie derivative (on vector of weight $\omega$) is also given in \cite{Berman:2012vc}\cite{Musaev2013}\cite{MusaevThesis}:
\be
\mathbb{L}_\Lambda V^{\alpha}=\Lambda^\beta\ptl_\beta V^\alpha-V^\beta\ptl_\beta \Lambda^\alpha+\fracs{1}{2}\gamma_a{}^{\alpha\beta}\gamma^a{}_{\gamma\delta}\ptl_\beta\Lambda^\gamma V^\delta.
\ee
Here $\gamma^a_{\alpha\beta}$ are $16\times 16$ chirally projected Gamma-matrices (off-diagonal components of 10D $32\times 32$ Gamma-matrices), $a=1,\dots,10$. They are symmetric and obey relations:
\be
\gamma^{(a}_{\alpha\beta}\gamma^{b),\beta\gamma}=\eta^{ab}\delta_\alpha^\gamma,\label{metric}
\ee
\be
\gamma^a_{(\alpha\beta}\gamma_{a,\gamma)\delta}=0.\label{Fierz}
\ee
$\eta^{ab}$ is the metric with (5,5) signature.

The $Z$-tensor in this case is:
\be
Z^{\alpha\beta}{}_{\gamma\delta}=\fracs{1}{2}\gamma_a{}^{\alpha\beta}\gamma^a{}_{\gamma\delta},
\ee
and the section condition is:
\be
\gamma_a{}^{\alpha\beta}\ptl_\alpha\otimes\ptl_\beta=0.\label{section6}
\ee

The E-brackets are in the usual form:
\be
[U,V]_E^\alpha=U^\beta\ptl_\beta V^\alpha+\fracs{1}{4}\gamma_a{}^{\alpha\beta}\gamma^a{}_{\gamma\delta}\ptl_\beta U^\gamma V^\delta-(U\leftrightarrow V).
\ee

The next tensor object is $B^a\in\mathfrak{B}(2\omega)$, which is in the vector representation of $SO(5,5)$. The isomorphism  from $B^{\alpha\beta}$ to $B_a$ is
\be
\bsp
(I_B B^{\alpha\beta})_a\ =\ &\fracs{1}{2}\gamma_{a,\alpha\beta}B^{\alpha\beta}\\[0.5ex]
(I_B^{-1}B_a)^{\alpha\beta}\ =\ &\fracs{1}{8}\gamma^{a,\alpha\beta}B_a.
\end{split}
\ee
The relevant formulas are:
\be
(A_1\bullet A_2)^a=\fracs{1}{2}\gamma_{\alpha\beta}^a A_1^\alpha A_2^\beta,
\ee
\be
(\wtl B)^\alpha=\gamma_a{}^{\alpha\beta}\ptl_\beta B^a.
\ee
Utilizing the Fierz identity (\ref{Fierz}), we can derive the form of generalized Lie derivative acting on $B^a$:
\be
\mathbb{L}_\Lambda B^a=\Lambda^\alpha\ptl_\alpha B^a+\fracs{1}{2}\gamma^a{}_{\alpha\gamma}\gamma_b{}^{\gamma\beta}B^b\ptl_\beta \Lambda^\alpha.
\ee

The next one $C_\alpha\in\mathfrak{C}(3\omega)$ lives in $\mathbf{\overline{16}}$ representation. The relevant rules are:
\be
(A\bullet B)_\alpha=\gamma^a{}_{\alpha\beta}B_a A^\beta,
\ee
\be
(\wtl C)^a=\fracs{1}{2}\gamma^{a,\alpha\beta}\ptl_\alpha C_\beta,
\ee
\be
\mathbb{L}_\Lambda C_\alpha=\Lambda^\beta\ptl_\beta C_\alpha+C_\beta\ptl_\alpha\Lambda^\beta-\fracs{1}{2}\gamma^a{}_{\alpha\beta}\gamma_a{}^{\gamma\delta}\ptl_\gamma\Lambda^\beta C_\delta+C_\alpha\ptl_\beta\Lambda^\beta.
\ee
The length of exceptional chain complex is $l=3$, and the specific weight $\omega=1/4$. 

The magic formula (\ref{magic}) in $d=6$ EFT only holds for $X=B\in\mathfrak{B}(2\omega)$:
\be
\bsp
(\mathbb{L}_\Lambda B)^a-\wtl(\Lambda\bullet B)^a-(\Lambda\bullet \wtl B)^a\ =\ &\Lambda^\alpha\ptl_\alpha B^a+\fracs{1}{2}\gamma^a{}_{\alpha\gamma}\gamma_b{}^{\gamma\beta}B^b\ptl_\beta \Lambda^\alpha\\[0.5ex]
&-\fracs{1}{2}\gamma^{a,\alpha\beta}\ptl_\alpha(\gamma^b{}_{\beta\gamma}B_b A^\gamma)-\fracs{1}{2}\gamma^a{}_{\alpha\beta}\Lambda^\alpha\gamma_b{}^{\beta\gamma}\ptl_\gamma B^b.
\end{split}
\ee
Using the definition (\ref{metric}) and relabeling the indices, the above quantity vanishes explicitly.

Another valid identity is the (\ref{AAAprop}), which actually means
\be
\gamma^a{}_{\alpha\beta}\gamma_{a,\gamma\delta}A_{(1}^\beta A_2^\gamma A_{3)}^\delta=0,
\ee
and directly follows from the Fierz identity (\ref{Fierz}).

Now we discuss how to solve the section condition. Actually the mathematics used in the current construction resembles the ``pure spinor formalism'', which is a way to quantize superstring in 10 dimensions\cite{Berkovits00}-\cite{Grassi01}. In that context, if a spinor $\lambda^\alpha$ satisfies the condition:
\be
\lambda^\alpha\gamma^a_{\alpha\beta}\lambda^\beta=0,
\ee
then it is called ``pure spinor''. One way to solve the equation is to decompose $|\lambda\rangle=\lambda^\alpha$ into GL(5) components:
\be
|\lambda\rangle=\lambda^+|0\rangle+\lambda_{i'j'}\Gamma^{i'j'}|0\rangle+\lambda^{i'}\epsilon_{i'j'k'l'm'}\Gamma^{m'l'k'j'}|0\rangle.
\ee
Here $i',j'\dots=1,\dots,5$ labels GL(5) fundamental index. $\Gamma^{i'}$ and $\Gamma_{i'}$ are linear combinations of the original $\Gamma^a$ matrices, so that the following relations hold:
\be
\{\Gamma_{i'},\Gamma^{j'}\}=\delta_{i'}^{j'}\ ,\ \{\Gamma_{i'},\Gamma_{j'}\}=0\ ,\ \{\Gamma^{i'},\Gamma^{j'}\}=0.
\ee
Hence the Majorana-Weyl representation $\mathbf{16}$ of SO(5,5) is decomposed into $\mathbf{16}=\mathbf{1}+\mathbf{\bar{5}}+\mathbf{10}$. Then the $\mathbf{\bar{5}}$ component $\lambda^{i'}$ can be written in terms of $\lambda^+$ and $\lambda_{i'j'}$, and the number of independent components of spinor $\lambda^\alpha$ is 11.

In \cite{Berman:2012vc}, it was shown that one can pick a pure spinor $\lambda^\alpha$, and impose the following linear condition:
\be
\lambda^\alpha(\gamma^{ab})_\alpha{}^\beta\ptl_\beta=0.
\ee
Then the section condition (\ref{section6}) can be solved, which finally gives the 6+5D M-theory solution.

Here we use a different argument to generate the M-theory and IIB solutions, which resembles the logic in \cite{E6}\cite{E7}. First we consider the embedding of GL(5) in SO(5,5). The branching rules of vector representation $\mathbf{10}$ and Majorana-Weyl representation $\mathbf{16}$ follows, where the subscripts denote the GL(1) weight:
\be
\bsp
&\mathbf{10}\rightarrow \mathbf{5}_{+2}+\mathbf{\bar{5}}_{-2}\\
&\mathbf{16}\rightarrow \mathbf{1}_{-5}+\mathbf{\bar{5}}_{+3}+\mathbf{10}_{-1}
\end{split}
\ee
Accordingly, we can then decompose the index $a=1,\dots,10$ and $\alpha=1,\dots,16$ into:
\be
\bsp
&a\rightarrow \{i\}, \{\bar{i}\} (i=1,\dots ,5)\\
&\alpha\rightarrow\{+\}\ ,\{\bar{i}\}\ ,\{ij\} (i,j=1\dots,5)
\end{split}
\ee
Under this decomposition, the tensor $\gamma_a^{\alpha\beta}$ is an GL(5) invariant tensor. This means that the component of $\gamma_a^{\alpha\beta}$ could be non-vanishing only if $\gamma_a^{\alpha\beta}$ carry zero GL(1) weight. Hence the non-vanishing components under the GL(5) decomposition are:
\be
(\gamma_i)^{+,\bar{j}}\ ,\ (\gamma_i)_{jk,lm}\ ,\ (\gamma^{\bar{i}})^{\bar{j},}{}_{kl}.
\ee
Then all the $(\gamma_a)^{\bar{i},\bar{j}}$ components vanish, which implies that choosing the $\bar{i}$ directions solves the section constraint (\ref{section6}).

For type IIB solution, we consider the embedding of SL(2)$\times$GL(4) in SO(5,5). The relevant branching rules are:
\be
\bsp
&\mathbf{10}\rightarrow (\mathbf{2},\mathbf{1})_{+1}+(\mathbf{2},\mathbf{1})_{-1}+(\mathbf{1},\mathbf{6})_0\\
&\mathbf{16}\rightarrow (\mathbf{1},\mathbf{4})_{+1}+(\mathbf{1},\mathbf{4})_{-1}+(\mathbf{2},\mathbf{\bar{4}})_0
\end{split}
\ee
Accordingly, we can decompose the index $a=1,\dots,10$ and $\alpha=1,\dots,16$ into:
\be
\bsp
&a\rightarrow \{\alpha'\}, \{\bar{\alpha'}\}, \{ij\}\ (\alpha'=1,2\ ;i,j=1,\dots ,4)\\
&\alpha\rightarrow\{i\},\{i'\},\{\bar{i}\alpha'\}\ (\alpha'=1,2\ ;i,i',\bar{i}=1\dots,4)
\end{split}
\ee
By the same arguments above, the non-vanishing components of the Gamma matrices are:
\be
(\gamma_{\alpha'})^{i',}{}_{\bar{j}\beta'}\ ,\ (\gamma_{\bar{\alpha'}})^{i,}{}_{\bar{j}\beta'}\ ,\ (\gamma_{ij})^{k,l'}\ ,\ (\gamma_{ij})_{\bar{k}\alpha',\bar{l}\beta'}.
\ee
This indicates that all the $(\gamma_a)^{i,j}$ and $(\gamma_a)^{i',j'}$ components vanish. Hence one can either choose $i$ or $i'$ to be the 4 interior directions giving the 6+4D IIB solution. These two choices are equivalent.

\section{Other groups and representations}

First we recollect the other existing symmetry groups in the literature. 

For $d=5$ EFT with group $G=E_{6(6)}$\cite{E6}, the vector $A^M$ is in the fundamental $\mathbf{27}$ representation. The length of exceptional chain complex is 2, and the other tensor object is $B_M\in\mathfrak{B}(2\omega)$, in the contragredient representation $\mathbf{\overline{27}}$. The invariant tensors of $E_{6(6)}$ includes rank-3 symmetric tensor $d^{MNK}$ and $d_{MNK}$. They are normalized by the following identity
\be
d^{MKL}d_{NKL}=\delta^M{}_N.
\ee
The $Z$-tensor is
\be
Z^{MN}{}_{KL}=10d_{KLP}d^{MNP},
\ee
and the section condition is
\be
d^{MNK}\ptl_M\otimes\ptl_N=0.
\ee

The identities involving $A^M\in\mathfrak{A}(\omega)$ and $B_M\in\mathfrak{B}(2\omega)$ are:
\be
\mathbb{L}_\Lambda A^M=\Lambda^N\ptl_N A^M-A^N\ptl_N\Lambda^M+10d_{NLP}d^{MKP}\ptl_K\Lambda^N A^L,
\ee
\be
\mathbb{L}_\Lambda B_M=\Lambda^N\ptl_N B_M+B_N\ptl_M\Lambda^N-10d_{MLP}d^{NKP}\ptl_K\Lambda^L A_N,
\ee
\be
(A_1\bullet A_2)_M=d_{MNK}A_1^N A_2^K,
\ee
\be
(\wtl B)^M=d^{MNK}\ptl_N B_K.
\ee
In this case none of the magic formulas in (\ref{magic}) holds, so the generalized Cartan calculus is not very interesting.

For the case of $d<5$, when $G=E_{7(7)}$ or $E_{8(8)}$, the expected length of exceptional exact sequence is $l=d-3<2$.  Fewer nice properties hold in these cases, hence we do not look into these cases. Nevertheless for the cases of $G=E_{6(6)}$, $E_{7(7)}$ or $E_{8(8)}$, because the exterior dimension is low, the higher-form fields (3-form gauge field or higher) do not appear in the construction of EFT. The generalized Cartan calculus is not relevant for these cases.

Another well-known example is the double field theory in Kaluza-Klein formalism\cite{HohmKK}. In that case the group is SO$(D,D)$, with $A^M$ always in the vector representation (note this is different from SO(5,5) EFT, where the 1-form gauge field is in Majorana-Weyl representation). The invariant tensor of SO$(D,D)$ are rank-2 symmetric tensor $\eta^{MN}$, $\eta_{MN}$, with relation
\be
\eta^{MK}\eta_{KN}=\delta^M{}_N.
\ee
The $Z$-tensor for double field theory is
\be
Z^{MN}{}_{PQ}=\eta^{MN}\eta_{PQ},
\ee
 and the section condition is the usual ``strong constraint'' in double field theory:
 \be
 \eta^{MN}\ptl_M\otimes\ptl_N=0.\label{strong}
 \ee
One can make a scalar using two vectors: $A^M A_M=\eta_{MN}A^M A^N$, and there are no other tensor objects in the theory. Hence this case is also trivial.

Apart from these trivial cases, one may wonder if there could be more possibilities. What happens when one just picks a group $G$ and some representation, and tries to construct a generalized Cartan calculus?

One possibility is to generalize the SL(3)$\times$SL(2) theory to group SL$(M)\times$SL$(N)$, $(M\geq N)$. The vector is in  $(\mathbf{M},\mathbf{N})$ representation, and we label the SL$(M)$ and SL$(N)$ indices by $i,j\dots$ and $\bar{i},\bar{j},\dots$ respectively. The $Z$-tensor for these theory are all in the same form, and it is not necessary to check the closure condition:
\be
Z^{i\bar{i},j\bar{j}}{}_{k\bar{k},l\bar{l}}=4\delta^{[i}_k\delta^{j]}_l\delta^{[\bar{i}}_{\bar{k}}\delta^{\bar{j}]}_{\bar{l}}.
\ee
The section condition is:
\be
\ptl_{[i[\bar{i}}\otimes\ptl_{j]\bar{j}]}=0.
\ee
Apparently the two maximal solutions correspond to $\ptl_{i1}\neq 0 (1\leq i\leq M)$ and $\ptl_{1j}\neq 0 (1\leq j\leq N)$. 

The other tensors can be written as contractions of multi-index tensors $X^{i_1 \bar{i}_1,i_2 \bar{i}_2,\dots i_n \bar{i}_n}$ with invariant tensors $\epsilon_{i_1 i_2\dots i_M}$ and $\epsilon_{\bar{i}_1\bar{i}_2\dots\bar{i}_N}$. When the least common multiple of $M$ and $N$ divides $n$, this tensor is a scalar (also a singlet). One can explicitly compute its weight\footnote{There is an alternative way to compute this weight, which is to write the $Z$-tensor in terms of projectors: $Z^{i\bar{i},j\bar{j}}{}_{k\bar{k},l\bar{l}}=\delta^i_k\delta^j_l\delta^{\bar{i}}_{\bar{k}}\delta^{\bar{j}}_{\bar{l}}-(N\mathbb{P}_{(M^2-1),1}+M\mathbb{P}_{1,(N^2-1)})^{i\bar{i}}{}_{l\bar{l}}{}^{j\bar{j}}{}_{k\bar{k}}+\frac{MN-M-N}{MN}\delta^i_l\delta^j_k\delta^{\bar{i}}_{\bar{l}}\delta^{\bar{j}}_{\bar{k}}$, and then read off $\omega=\frac{MN-M-N}{MN}$. This is consistent with the method which computes the transformation rule of a multiple-index tensor.}, which is
\be
n\omega=\frac{n(MN-M-N)}{MN}.
\ee
To get a scalar with weight 1, the only possible triplets $(M,N,n)$ are:
\be
(M,N,n)=(3,2,6)\ ,\ (3,3,3)\ ,\ (4,2,4) .
\ee

Hence we conclude that the only possible EFTs of this particular type carry group SL(3)$\times$SL(2) or SL(3)$\times$SL(3) or SL(4)$\times$SL(2). From the computation of weight, one can obtain the dimension of these theories. SL(3)$\times$SL(2) is the usual $d=8$ EFT, in 8+6D, and the two solutions to the section condtion give 8+3D and 8+2D theories. The SL(3)$\times $SL(3) EFT lives in 5+9D, and the two solutions both result in 5+3D theories. The SL(4)$\times$SL(2) EFT lives in 6+8D, and the two solutions give 6+4D and 6+2D theories. 

Actually SL(3)$\times$SL(3) and SL(4)$\times$SL(2) are the subalgebra of $E_6$ and SO(10) respectively. Hence they may be treated as some truncated version of $d=5$ and $d=6$ EFT.

Another possible extension is to consider group $G=$SL$(M)$, and the vector fields are in the rank-2 anti-symmetric tensor representation(similar construction is discussed in \cite{Park:2014una}\cite{Strickland-Constable:2013xta}). The interior directions are labelled by $mn$, and there are in total $M(M-1)/2$ of them. The form of $Z$-tensor and section condition are similar to the SL(5) case:
\be
Z^{ij,kl}{}_{pq,rs}=\fracs{1}{4}\fracs{1}{(M-4)!}\epsilon^{ijklm_1\dots m_{M-4}}\epsilon_{pqrsm_1\dots m_{M-4}},
\ee
\be
\epsilon^{ijklm_1\dots m_{M-4}}\ptl_{ij}\otimes\ptl_{kl}=0.
\ee
The other tensor objects are obtained by contracting multi-index tensors with $\epsilon_{i_1 i_2\dots i_M}$. The singlet in this theory is given by
\be
\epsilon_{i_1 i_2\dots i_M}\epsilon_{j_1 j_2\dots j_M}X^{i_1 j_1,i_2 j_2,\dots,i_M j_M}.
\ee
Its weight turns out to be always 1, so that any value of $M$ is acceptable. The theory lives in $M+2$ exterior dimensions and $M(M-1)/2$ interior dimensions. The distinct solutions to section condition are 
\be
\ptl_{12},\ptl_{13},\dots,\ptl_{1M}\neq 0
\ee
and
\be
\ptl_{12},\ptl_{13},\ptl_{23}\neq 0.
\ee
Hence the theory after solving section condition is in $2M+1$ or $M+5$ dimensions.

 The physical meaning of these new possiblities is not clear. It is also unknown whether full EFTs with these groups exist. If they do exist, then it is also not clear whether these theories admit supersymmetric extensions. Also, we do not give a systematic classfication of all the allowed groups and representations. We leave these questions to future research. 

\section{Conclusion}

In this paper we explicitly constructed the generalized Cartan calculus for $d=9$, $d=7$ and $d=6$ EFT. The nice properties  guarantee that in $d=7$ and $d=6$ EFT, one can construct a gauge covariant 4-form field strength and a 3-form field strength respectively. In $d=9$ EFT, one can at least construct a gauge covariant 5-form field strength, which should is useful for constructing the EFT action. With the tensor hierarchy, one is able to construct the action following the logic in \cite{E6}\cite{8D}. Demanding that the action is invariant under both interior gauge transformations ($\delta_\Lambda\ ,\  \delta_\Xi$ etc.) and exterior diffeomorphisms will fix the form of each term and their relative coefficient in the action. In even dimensions ($d=4,6,8$), one should impose some duality conditions on field strengths, which is reminiscent of the self-duality condition in IIB supergravity, hence the action is a pseudo-action. One should also be able to write down the supersymmetric extension of these EFTs, following the procedure in \cite{SUSYE7}\cite{SUSYE6}.

Also, if one explicitly writes out all the indices for a specific dimension, say $d=6$, then the relevant formulas of the tensor hierarchy in Section 3 resemble the corresponding ones in gauged supergravity\cite{Bergshoeff08}. One can use a generalized Scherk-Schwarz ansatz in the duality manifest theory to reproduce gauged supergravity\cite{Hohm:2014qga}, and the formalism developed here may help to simplify some computations in gauged supergravity context.

We also found here that the group used in generalized Cartan calculus is not restricted to the exceptional series. It would be interesting if a full theory with exterior diffeomorphism invariance could be constructed. For example, it may be also possible to incorporate other theories into this framework, for example, half-maximal gauged supergravity. 

\acknowledgments

The author thanks Olaf Hohm and Barton Zwiebach for useful discussions and comments on the manuscripts. The author also thanks David Berman and Henning Samtleben for discussions.

This work is supported by the U.S. Department of Energy under grant Contract Number  DE-SC00012567.

\providecommand{\href}[2]{#2}\begingroup\raggedright

\end{document}